\documentclass[onecolumn]{aa}  

\usepackage{graphicx}  
\usepackage[varg]{txfonts}
\usepackage{hyperref}
\hypersetup{
    colorlinks=true,
    linkcolor=blue,
    filecolor=magenta,      
    urlcolor=cyan,
   citecolor=blue,
}
\defcitealias{2020kri}{Paper I}
\begin{document} 

\definecolor{amber}{rgb}{1.0, 0.75, 0.0}
\definecolor{pumpkin}{rgb}{1.0, 0.46, 0.09}
\definecolor{cadmiumred}{rgb}{0.89, 0.0, 0.13}
\definecolor{azure}{rgb}{0.0, 0.5, 1.0}
\definecolor{darkorchid}{rgb}{0.6, 0.2, 0.8}
\definecolor{phthaloblue}{rgb}{0.0, 0.06, 0.54}
\definecolor{britishracinggreen}{rgb}{0.0, 0.26, 0.15}
\definecolor{byzantine}{rgb}{0.74, 0.2, 0.64}
\definecolor{verdigris}{rgb}{0.47, 0.32, 0.66}
\definecolor{verdigris}{rgb}{0.26, 0.7, 0.68}

   \title{Thermal and kinetic coronal rain diagnostics with \ion{Mg}{II} h \& k lines }

   \author{M. Kriginsky\inst{1,2} \and R. Oliver\inst{1,2} }

   \institute{Departament de F\'\i sica, Universitat de les Illes Balears, E-07122 Palma de Mallorca, Spain
         \and
          Institute of Applied Computing \& Community Code (IAC3), UIB, Spain
               }

   \date{Received ; accepted }

 
  \abstract
   {}
   {The aim of this work is to assess the temperature and velocity diagnostics of coronal rain clumps, as observed in the lines formed in the upper chromosphere.}
   {We performed the temperature diagnostics using inversions of data from nine spectroscopic observations obtained with the IRIS spectrograph in the \ion{Mg}{II} h \& k lines. The sensitivity to the temperature of the emission peaks of these lines was exploited  to determine the temperature of the coronal rain plasma using inversions of the spectroscopic profiles. Additional relationships between different spectral features of these lines, derived from the use of 3D radiative transfer line synthesis applied to simulations, were employed in order to derive the line-of-sight (LoS) velocities in different parts of the coronal rain plasma. }
   {For the first time, spectroscopic inversions of coronal rain were successfully performed. Temperatures derived from the inversions yield coronal rain clump temperatures at the formation height of the emission peaks of the \ion{Mg}{II} h \& k lines in the range between 5000 and 7000~K. This narrow range of values remains consistent among all the different observations used in this work. We obtained LoS velocities of up to 40 km $\mathrm{s}^{-1}$, which are consistent with the motion of the plasma being mostly constrained to the plane of the sky, as the coronal rain was mostly detected shortly after its formation and the observations took place in the disc. Furthermore, velocity diagnostics led to the detection of larger velocities at higher layers of the coronal rain plasma in some cases. This increased velocity seems to indicate that at some point (at least) during the fall of coronal rain clumps towards the chromosphere, the material in the upper part of the coronal rain plasma is falling with greater velocity than the material below it. The conditions of the temperature and density of the coronal rain plasma where the \ion{Mg}{II} h line forms appear to be slightly different that those of the  \ion{Mg}{II} k line, with the former found at slightly colder and denser parts of the plasma.  }
   {}


   \titlerunning{}
   \authorrunning{M. Kriginsky et al.}
   \maketitle
   
%

\section{Introduction}

Numerical simulations where the heating is concentrated chiefly towards the footpoints of coronal loops can lead to a phenomenon known as coronal rain \citep[e.g.][]{1991ApJ...378..372A,2003A&A...411..605M,2004A&A...424..289M,2013ApJ...773...94M}. Footpoint heating leads the chromospheric plasma to `evaporate' and rise along the loop, ultimately reaching coronal temperatures and therefore filling it with coronal-temperature material. As the heating is predominantly concentrated at the footpoints, the material near the coronal loop apex is unable to maintain its thermodynamic state due to the radiative losses. These increase as the density of the plasma increases and its temperature diminishes, a dynamics that can sometimes lead to a runaway cooling process if the temperature falls below a certain threshold, usually at $\sim$$10^6$~K. In this situation, the plasma will continue to cool down to transition region and chromospheric temperatures, while at the same time increasing its density. Such dense condensation is out of the mechanical equilibrium and flows down along the loop in the form of coronal rain.

The coronal rain material consists of a denser and colder core of chromospheric temperature surrounded by a warm envelope, whose temperature transitions from that of the clump's core to coronal values \citep{2022ApJ...926L..29A}. This temperature inhomogeneity allows for the plasma to be visible in a myriad of spectral lines that sample coronal and chromospheric temperatures. Coronal rain appears in these spectral lines as clump-like concentrations of material falling along the magnetic field of coronal loops with velocities of up to $100-150$~$\mathrm{km}\mathrm{s}^{-1}$ \citep{2010ApJ...716..154A,2012ApJ...745..152A,2015ApJ...806...81A,2020A&A...633A..11F,2021A&A...650A..71K}. Spectroscopic observations of coronal rain in chromospheric lines, such as the \ion{Ca}{II} K and H$\alpha$ lines, with high-resolution instruments such as the Solar Optical Telescope on board of Hinode \citep{2008SoPh..249..167T} and the CRISP imaging spectropolarimeter \citep{Scharmer_2008} at the Swedish 1-meter solar telescope \citep{sst} have allowed for the determination of multiple properties of coronal rain clumps, such as their temperature and downfall velocities. The use of off-limb observations in chromospheric lines forming in the optically thin regime allowed for several estimates of the average temperature of coronal rain clumps \citep{2012ApJ...745..152A,2015ApJ...806...81A,2020A&A...633A..11F}. These studies have found temperatures of coronal rain plasma in the range between 5000~K up to values beyond 30 000~K.

\begin{table*}[h!]
         \renewcommand*{\arraystretch}{1.5}
        \centering
        \caption{ Description of the datasets used in this work. The modes s.a.s. and r.s. stand for sit-and-stare and raster scans, respectively.}
        \label{table:Table1}
        \begin{tabular}{ccccccccc}
        
            \hline \hline
            
            
            Number&Date&Start$-$end &OBSID&Solar X, Y &Exp.&Raster FOV& AR No.&Mode\\
            &&time (UT)&&&time [s]&&&\\
            \hline
            \textcolor{black}{1}&\textcolor{black}{19 March 2014} &\textcolor{black}{04:33$-$05:12} &\textcolor{black}{3800257403}&\textcolor{black}{ -28\arcsec, 336\arcsec}&\textcolor{black}{5}&\textcolor{black}{0.33\arcsec,119\arcsec}&\textcolor{black}{12007}&\textcolor{black}{s.a.s.}\\
            
            \textcolor{black}{2}&\textcolor{black}{19 March 2014} &\textcolor{black}{11:56$-$13:37} &\textcolor{black}{3800257403}& \textcolor{black}{34\arcsec, 328\arcsec}&\textcolor{black}{5}&\textcolor{black}{0.33\arcsec,119\arcsec}&\textcolor{black}{12007}&\textcolor{black}{s.a.s.}\\
            
            \textcolor{black}{3}&\textcolor{black}{19 March 2014} &\textcolor{black}{13:46$-$15:57} &\textcolor{black}{3800257403}& \textcolor{black}{62\arcsec, 330\arcsec}&\textcolor{black}{5}&\textcolor{black}{0.33\arcsec,119\arcsec}&\textcolor{black}{12007}&\textcolor{black}{s.a.s.}\\
            
            \textcolor{black}{4}&\textcolor{black}{20 March 2014} &\textcolor{black}{18:15$-$19:33} &\textcolor{black}{3800281286}&\textcolor{black}{328\arcsec, 328\arcsec}&\textcolor{black}{15}&\textcolor{black}{15\arcsec,119\arcsec}&\textcolor{black}{12007}&\textcolor{black}{r.s.}\\
            
            \textcolor{black}{5}&\textcolor{black}{20 March 2014} &\textcolor{black}{19:54$-$21:12} &\textcolor{black}{3800281286}& \textcolor{black}{353\arcsec, 318\arcsec}&\textcolor{black}{15}&\textcolor{black}{15\arcsec,119\arcsec}&\textcolor{black}{12007}&\textcolor{black}{r.s.}\\
            
            \textcolor{black}{6}&\textcolor{black}{20 March 2014} &\textcolor{black}{21:31$-$22:49} &\textcolor{black}{3800281286}& \textcolor{black}{365\arcsec, 325\arcsec}&\textcolor{black}{15}&\textcolor{black}{15\arcsec,119\arcsec}&\textcolor{black}{12007}&\textcolor{black}{r.s.}\\
            
            \textcolor{black}{7}&\textcolor{black}{13/14 May 2014} &\textcolor{black}{23:09$-$03:21} &\textcolor{black}{3820609153}& \textcolor{black}{422\arcsec, 132\arcsec}&\textcolor{black}{9}&\textcolor{black}{0.33\arcsec,119\arcsec}&\textcolor{black}{12056}&\textcolor{black}{s.a.s.}\\
            
            \textcolor{black}{8}&\textcolor{black}{10 July 2015} &\textcolor{black}{10:39$-$14:40} &\textcolor{black}{3660106017}& \textcolor{black}{251\arcsec, 189\arcsec}&\textcolor{black}{5}&\textcolor{black}{1\arcsec,119\arcsec}&\textcolor{black}{12381}&\textcolor{black}{r.s.}\\
            
            \textcolor{black}{9}&\textcolor{black}{24 December 2015} &\textcolor{black}{00:39$-$01:37} &\textcolor{black}{3620106014}& \textcolor{black}{-834\arcsec, 114\arcsec}&\textcolor{black}{5}&\textcolor{black}{2\arcsec,175\arcsec}&\textcolor{black}{12472}&\textcolor{black}{r.s.}\\\hline

        \end{tabular}
\end{table*}

The focus of this work is to use another diagnostic technique that still has not been exploited in coronal rain observations: spectroscopic inversions. When the coronal rain material reaches chromospheric temperatures and can be detected on the disc with enough signal in a spectral line whose formation properties are strongly correlated  with those of the plasma constituting the coronal rain clumps, the use of inversion codes might serve as an alternative diagnostic tool. textcolor{black}{The diagnostic of the properties of the coronal rain plasma is most easily done in off-limb observations. However, if the line under study has enough opacity in coronal rain material for it to be formed under optically thick conditions, it could also potentially be used in on-disc observations.} The \ion{Mg}{II} h \& k lines, which form in the upper chromosphere and possess stronger opacity than the more extensively used \ion{Ca}{II} H \& K lines are possibly among the most promising candidates.

The launch of the Interface Region Imaging Spectrograph \citep[IRIS,][]{2014SoPh..289.2733D} mission has made systematic observations of the chromosphere and transition region broadly available, including spectra from lines that are inaccessible from ground-based solar observatories. The near-ultraviolet (near-UV) spectral range of IRIS includes spectroscopic data from both the  \ion{Mg}{II} h \& k lines and the secondary triplet of \ion{Mg}{II} located in their vicinity. With IRIS completing a decade of observations, the online database now includes a multitude of coronal rain observations, most easily seen off-limb. 

Before and after the launch of IRIS, a series of publications have been devoted to the study of the different spectral regions and lines accessible to its spectrograph. In particular, the works of \citet{2013ApJ...772...89L,2013ApJ...772...90L} focused on the formation properties of the \ion{Mg}{II} h \& k lines and the connection of such properties to the physical state of the plasma. This was done with the use of synthesis of the spectral profiles computed numerically using as the input atmosphere a simulation snapshot from the Bifrost code \citep{2011A&A...531A.154G}. The aforementioned works found a strong relationship between the temperature at the formation height of the emission peaks of the \ion{Mg}{II} h \& k lines and their intensity; therefore, their inclusion in inversions could lead to a better constraint of the temperature at their formation height. This temperature sensitivity has been exploited in a number of works \citep{2016ApJ...830L..30D,2018A&A...620A.124D,2019A&A...627A.101V}. In addition to constraining the temperature of the plasma, \citet{2013ApJ...772...90L} showed that, at least for an atmosphere arising from a simulation snapshot, there are some constraints about the vertical velocity of the plasma that can be obtained by combining different spectral features of the \ion{Mg}{II} h \& k lines.

The goal of the present work is to exploit the diagnostic capabilities of the \ion{Mg}{II} h \& k lines in order to infer properties about the coronal rain clumps forming in active regions observed with the IRIS spectrograph. This work is organised as follows. Section~\ref{sec:obs} details the different observations used and their calibration, with the inversion scheme and data analysis of spectral features detailed in Sect.~\ref{sec:Data}. The results are discussed in Sect.~\ref{sec:results} and our conclusions are presented in Sect.~\ref{sec:conc}

 \section{Observations} \label{sec:obs}

  Coronal rain is nearly always present in active regions \citep{2023ApJ...950..171S}. However, its detection on the disc is not straightforward. The presence of a myriad of different phenomena in active regions hinders the possibility of detecting coronal rain events automatically. In this work, a comprehensive manual inspection of the IRIS public data archive\footnote{\url{https://iris.lmsal.com/search/}} was undertaken in order to detect on-disc coronal rain events that crossed the slit position at any given point in their evolution. 

  The study of the IRIS observations in coordination with data in the 30.4 nm passband from the Atmospheric Imaging Assembly \citep[AIA,][]{2012SoPh..275...41B,2012SoPh..275...17L}, on board the Solar Dynamics Observatory \citep[SDO,][]{2012SoPh..275....3P},  demonstrated that the easiest way to detect coronal rain clumps in active regions is when the slit is positioned in the neighbourhood of the bright plage areas and the coronal loop hosting the coronal rain event is crossing the field of view of this neighbourhood. This situation allows for the easiest detection of the presence of coronal rain in the slitjaw images of IRIS as well as in the spectra of the \ion{Mg}{II} h \& k lines. The presence of coronal rain in the field of view corresponds to a diminution of over 40 \% of the intensity compared to the background. This diminution is much smaller or nonexistent in the case of coronal rain clumps crossing a quiet Sun area. Therefore, it becomes nearly impossible to unambiguously detect the motion and position of coronal rain in the slitjaw  in such a scenario.

The data used in this work consisted of nine observations undertaken with the IRIS spacecraft between 2014 and 2015. All observations targeted active regions. In each one, coronal rain is seen crossing the position of the slit, with the slit either in a sit-and-stare mode or performing raster scans. A detailed description of each observation is provided in Table~\ref{table:Table1}. The calibration of the data was done in line with the reduction methods detailed in \citet[][]{2018SoPh..293..149W}. Radiometric calibration of the raster data was carried out using version four of the calibration files employed by the
\textit{iris\_get\_response} routine in SolarSoft \citep[SSW;][]{1998SoPh..182..497F}.

\begin{figure*}[p!]
 \centering
 \includegraphics[width=17cm]{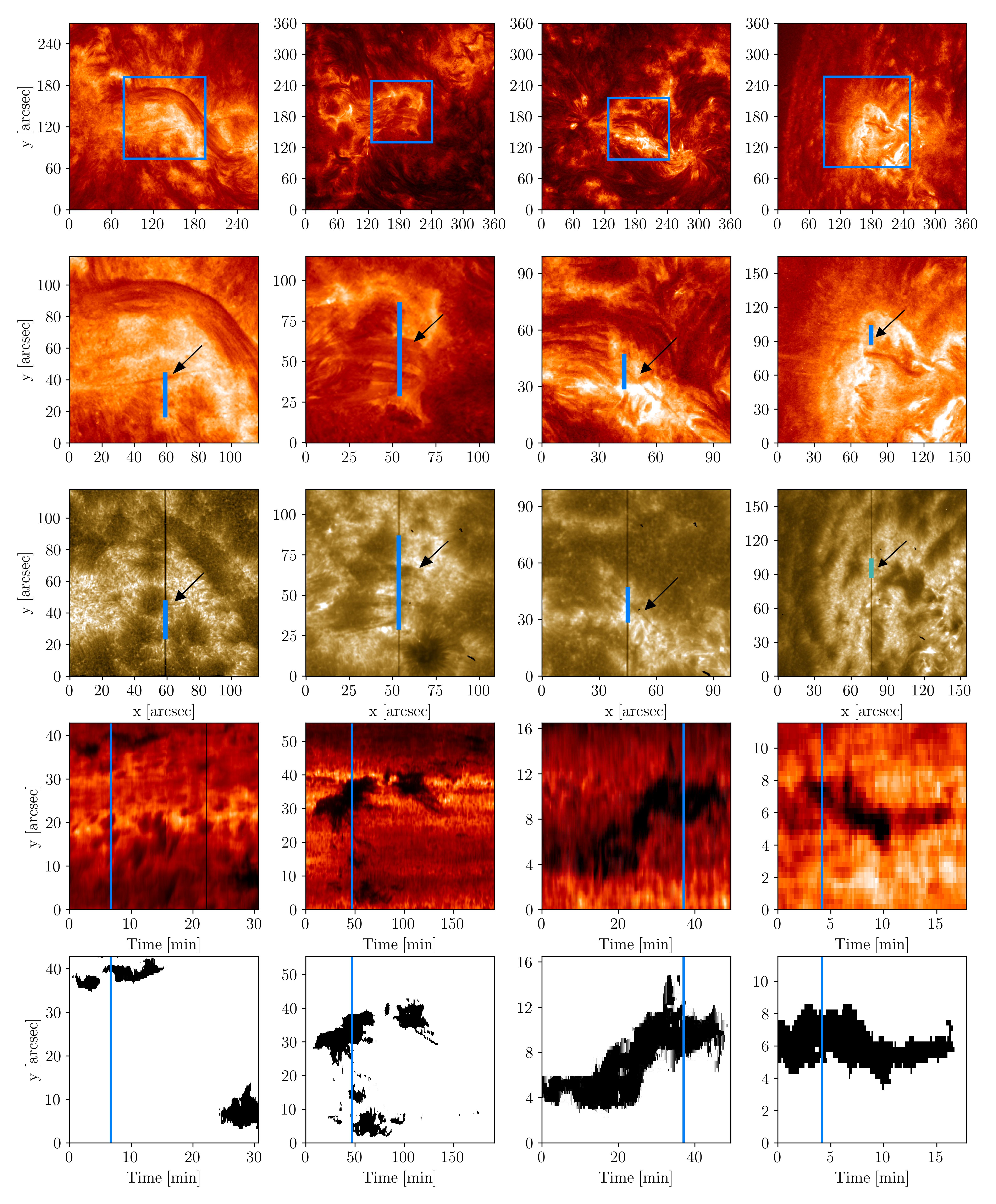}
 \caption{Observations used in this study. Each column corresponds to one of four of the observations listed in Table~\ref{table:Table1}, namely: observations 1, 7, 8, and 9. The first row displays the field of view of the IRIS slitjaw imager as a coloured rectangle within a co-aligned image in the 30.4 nm passband obtainedw with SDO/AIA. The second row panels display a zoomed view into the field of view inside the rectangles of the first row panels. The corresponding slitjaw images in the 279.6 nm passband are displayed in the third row panels. The thick vertical lines in each panel of the second and third rows outline the spatial portion of the slit used in the fourth row panels. Arrows indicate the location of coronal rain as it crosses the slit. The fourth row panels display the intensity in the core of the \ion{Mg}{II} h line as a function of time. The vertical coloured lines in each panel of the fourth row provides the time instant used for the images in the top three corresponding panels. Finally, the last row shows a mask of the fourth row panels for each observation, with the pixels used for the inversions in each case shown in black. The areas  coloured white  were not used in the inversions. Two accompanying movies display the coronal rain formation and its crossing of the slit position for observations 7 and 8. The two movies have been trimmed to show the time when coronal rain is present.}
 \label{Figure:2}%
\end{figure*}

Referring to the labels in Table~\ref{table:Table1}, observations 1-6 correspond to a tracking campaign of active region 12007. The remainder of the observations correspond to single observations of different active regions where coronal rain was visibly crossing the slit. In observations 1-3 and observation 7, the scan mode was sit-and-stare; whereas for observations 4 -6 a sparse (1\arcsec steps) raster scan was undertaken and for observations 8-9 dense (0.33\arcsec steps) and coarse (2\arcsec steps) scans were performed, respectively. Figure~\ref{Figure:2} provides a context of the observations for each of the different active regions listed in Table~\ref{table:Table1}. For brevity and clarity, only one observation corresponding to active region number 12007 is shown, as these observations are largely similar to each other, differing only with respect to the particular coronal loop that hosts the coronal rain event during the observation. Coronal rain material in the shape of long strands can be clearly seen in the AIA 30.4 nm images (first and second rows of Fig.~\ref{Figure:2}). Detecting this material in the 279.6 nm passband is much harder (third row of Fig.~\ref{Figure:2}).

\section{Methods}\label{sec:Data}

 \subsection{Inversions}

 The inversion of the IRIS data was made with the MPI-parallel Stockholm inversion code \citep[STiC,][]{2016ApJ...830L..30D,2019A&A...623A..74D}, which is a regularised Levenberg-Marquardt code. It uses an optimised version of the RH code \citep[][]{2001ApJ...557..389U} to solve the atom population densities assuming statistical equililibrium and plane-parallel geometry for multi-atom inversions of multiple spectral lines assuming non-local thermodynamic equilibrium (non-LTE). It also allows for the inclusion of partial redistribution effects of scattered photons \citep[][]{2012A&A...543A.109L}. The radiative transport equation is solved using cubic Bezier solvers \citep[][]{2013ApJ...764...33D}. The inversion engine of STiC includes an equation of state extracted from the Spectroscopy Made Easy code \citep[SME;][]{2017A&A...597A..16P}. The electron densities were derived with the assumption of non-LTE hydrogen ionisation by solving the statistical equilibrium equations imposing charge conservation \citep[][]{2007A&A...473..625L}. As a 1.5D plane parallel code, STiC cannot account for some 3D radiative transfer effects that could be important for the inversion of the \ion{Mg}{II} h \& k lines \citep[][]{2013ApJ...772...89L,2013ApJ...772...90L,2015ApJ...806...14P}.

 A model atom of \ion{Mg}{II} consisting of 11 atomic levels including the ground level of \ion{Mg}{III} was used. In addition, a photospheric \ion{Ni}{I} line at 281.4 nm was used in order to have some constraint about the state of the photosphere, which was treated under the assumption of LTE. In order to save computation time given the large number of inversions performed, only the region around the cores of the \ion{Mg}{II} h \& k lines was used for the inversions, using the same spectral regions as those used by \citet[][in particular, their Fig.~4]{2023A&A...672A..89K}.

  A regular depth grid of 36 points covering the range $\log \tau_{500} = [-7,0]$ (where $\tau_{500}$ is the optical depth of the 500~nm radiation) was used in the inversions. The initial temperature depth profile was interpolated from a FAL-C \citep[][]{1985cdm..proc...67A,1993ApJ...406..319F} atmosphere and used as an initial guess for the inversions. Given the different view points in the observations, the initial guess for the velocity along the line of sight (LoS) was modified in order to provide a suitable guess for each observation. The parameters that were actively inverted were the temperature, $T$, the LoS component of the velocity, $v_{\mathrm{LOS}}$, and the microturbulent velocity, $v_{\mathrm{turb}}$. The node distribution was not the same for every observation. However, the same philosophy was followed, namely an increasing number of nodes for each cycle. Most of the inversions were performed in no more than three inversion cycles, except for observation 9, which required up to five inversion cycles in order to reach a reasonable fit to the observations. This is probably due to to the point of view of the observation being much closer to the limb than the other observations.

  Given the large amount of pixels in the data from all the observations, only the pixels containing coronal rain were inverted. They were selected taking advantage of the large intensity difference between pixels containing coronal rain and background plage pixels through an intensity masking procedure (see bottom row of Fig~\ref{Figure:2}). Additionally, in order to reduce the number of wavelength points used in the inversions, a region centred at the rest wavelengths of the \ion{Mg}{II} h \& k lines with a total width of 0.2 nm was used. For the two \ion{Mg}{II} triplet lines between the h \& k lines, an additional region with a width of 0.1 nm was selected, centred at the 279.9 nm wavelength.

 \subsection{Velocity diagnostics}\label{sect:43}

 Macroscopic velocity fields can be inferred from the Doppler displacement of a spectral line. This procedure has been extensively used before with spectra obtained in chromospheric lines, such as the \ion{Ca}{II} H line \citep{2010ApJ...716..154A} or the H$\alpha$ line,  to determine the typical velocities of coronal rain clumps \citep{2010ApJ...716..154A,2012ApJ...745..152A,2015ApJ...806...81A,2020A&A...633A..11F,2021A&A...650A..71K}. The results of these works are in great agreement with each other, with clump total velocities of up to 200 km$\mathrm{s^{-1}}$  obtained.

 The spectra in the \ion{Mg}{II} h \& k lines also allows for a similar study. In fact, there are several additional diagnostics that are possible by using both lines individually and by combining them. Using spectra computed with the Multi3D radiative transfer code \citep{2009ASPC..415...87L} with data from a simulation snapshot performed with the Bifrost code \citep{2011A&A...531A.154G}, \citet{2013ApJ...772...90L} performed a series of tests using several spectral features of these lines in order to assess what information can be extracted from them. The diagnostics used in this work are described below (see \citet{2013ApJ...772...90L} for the complete discussion and for the definition of the nomenclature used below): 

 \begin{itemize}
     \item With respect to the h3 and k3 Doppler shifts: The Doppler displacements of the line core position of the \ion{Mg}{II} h \& k lines are found to be strongly correlated with the velocity along the LoS at the height of formation of the line cores. This value is hereafter referred to as either $v_3$ when grouping together results for both h \& k lines or as $v_{h_3}$ and $v_{k_3}$ when specifying the line becomes relevant.

     \item Difference between h3 and k3: The core of the k line was found to be formed a few hundred km below the transition region, with the core of the h line forming at slightly lower heights. For small differences (smaller than 2 km$\mathrm{s^{-1}}$) between the velocities at the formation heights of h3 and k3, there is a strong correlation between such differences and the difference between the velocities measured from Doppler shifts at h3 and k3. 
     
     \item Average peak Doppler shift: A correlation between the average Doppler shift of the red and blue emission peaks of both lines was found to be tightly related to the average vertical velocity at the height of formation of the peaks. This velocity value is referred to here as $v_2$ as a general term for either the h or k lines and explicitly as $v_{h_2}$ or $v_{k_2}$ when the value for each line is referenced.

 \end{itemize}

 textcolor{black}{Access to these diagnostics  makes it possible to estimate the vertical
velocities and their differences inside the coronal clump plasma.}
 
\begin{figure}[h!]
 \centering
 \includegraphics[width=8cm]{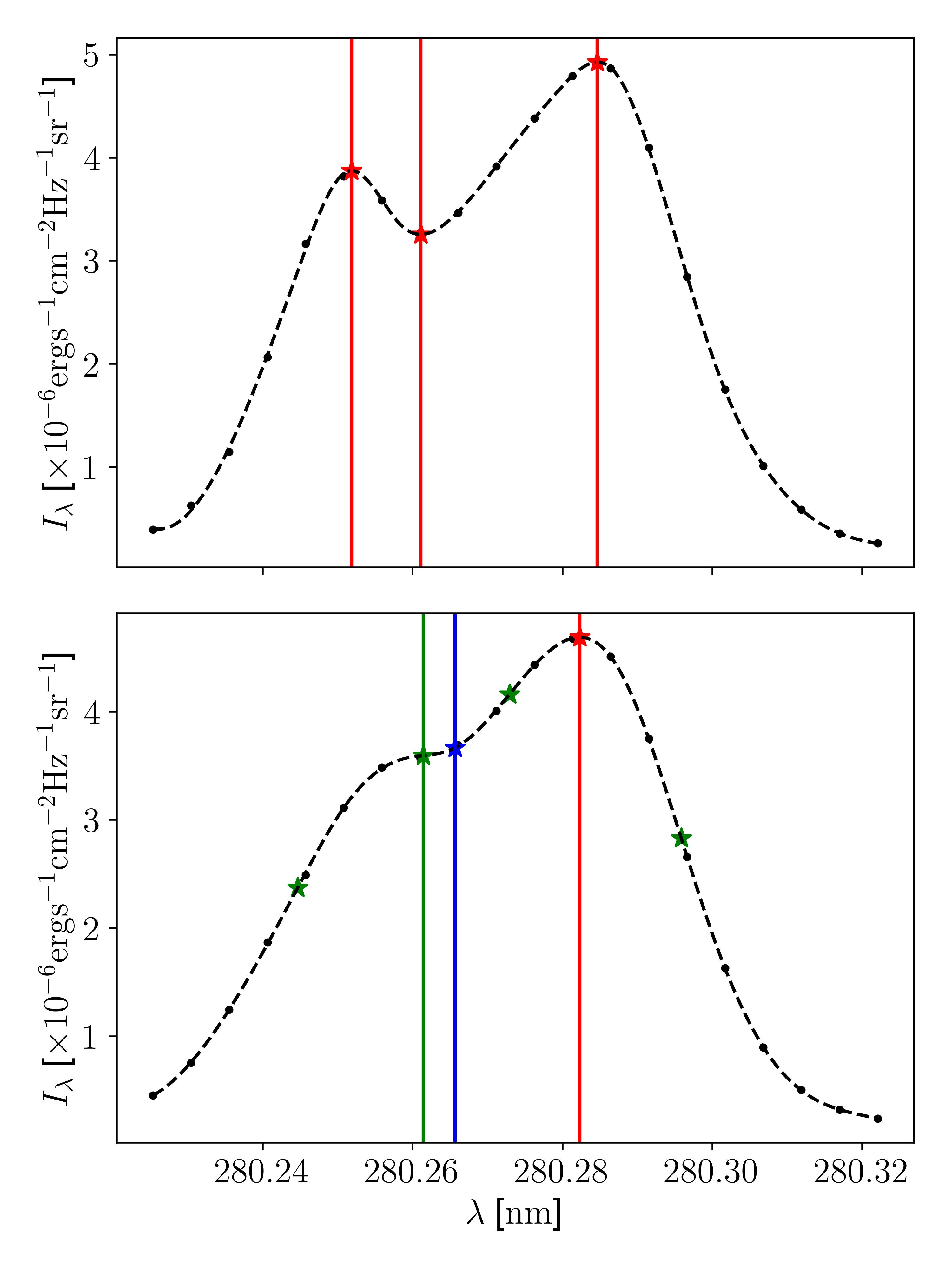}
 \caption{Spectral features, featuring a typical \ion{Mg}{II} h line profile with two emission peaks and a line core minimum (shown in the top panel) and a single-peaked profile (shown in the bottom panel). The positions of the extremes of the fitted profile are marked with a red star, with the vertical red lines denoting the retrieved h3 and h2 features. For the single-peaked profile, the position of the only extreme is also marked with a red star, while the extremes of the first and second derivatives are marked with green and blue stars, respectively. The vertical lines also identify the retrieved h3 and h2 features. The black dashed curves in both panels represent the fitted curve to the spectral profile, with the black dots representing observed spectral points. }
 \label{Figure:4}%
\end{figure}

\subsection{Identification of spectral features}\label{sect:44}
In order to make use of the diagnostics given above, the different spectral features (emission peaks and central reversal) had to be determined for each pixel. This was done by means of fitting the region around the line cores of the \ion{Mg}{II} h \& k lines with a sum of splines. There are mainly four types of \ion{Mg}{II} h \& k line core profiles, namely:\ the typically double-peaked profile with a central reversal (top panel of Fig~\ref{Figure:4}), a single-peaked profile with a secondary lobe (bottom panel of Fig~\ref{Figure:4}), a single-peaked profile without a secondary lobe, and profiles with more than two emission peaks. In order to minimise the uncertainties when determining the position of the h2, h3, k2, and k3 spectral features, only the first two types of profiles were used, and the rest were discarded. This represented a loss of about 20\% of all coronal rain pixels.

For the regular, double-peaked profiles, the detection algorithm is straightforward: the extremes of the first derivative of the fitted curve are determined, with the two maxima and the minimum between them extracted. Profiles with one maximum and a secondary lobe required the use of the second and third derivatives. The position of either the red or blue emission peak was determined as the zero crossing position of the first derivative. These profiles are characterised by four points where the second derivative is zero. If the profile maximum is in the red (blue) side, then the second point with null second derivative to its right (left) was taken as the position of the remaining emission peak. The line centre was taken to be the position of the zero crossing of the third derivative in between the two emission peaks (see Fig~\ref{Figure:4}).

\begin{figure*}[h!]
 \centering
 \includegraphics[width=16cm,height=5cm]{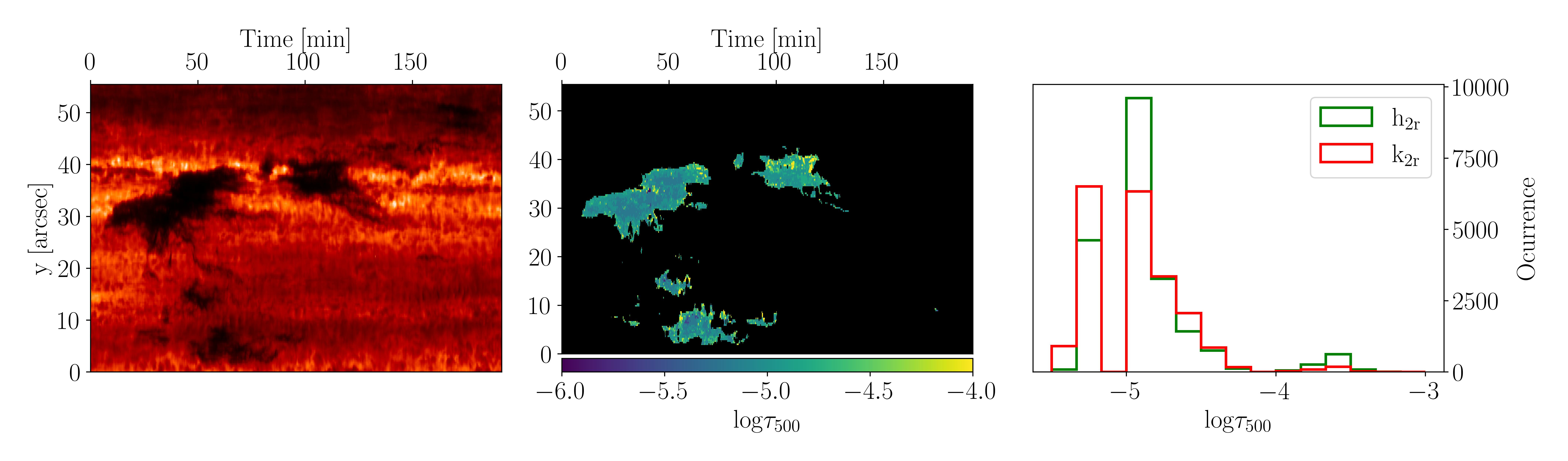}
 \includegraphics[width=16cm,height=5cm]{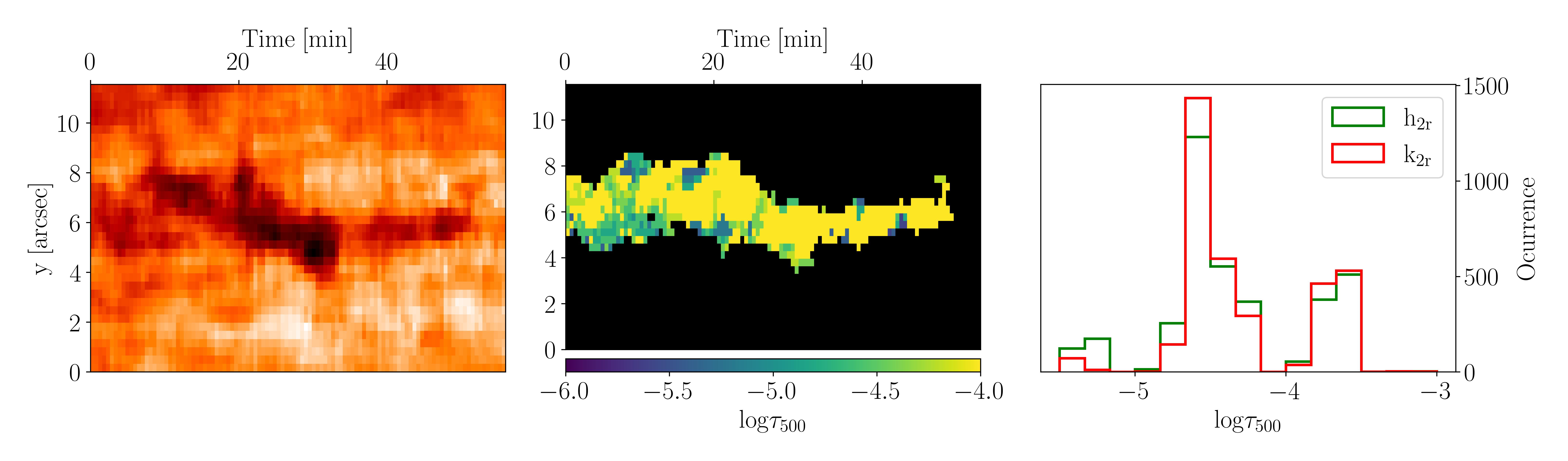}
 \caption{Response functions. Left columns display the line core intensity in the \ion{Mg}{II} k line as a function of time, with the top one referring to observation 7, and the bottom to observation 9. The middle panels show the optical depth of maximum response to temperature in these observations for the $h_{2r}$ feature. The right columns show the distributions of maximum response to temperature for each observation, in each case, for both $h_{2r}$ and $k_{2r}$ features. }
 \label{Figure:7}%
\end{figure*}

\section{Results} \label{sec:results}

\subsection{Response functions}
The optical depth of formation of each wavelength was approximated by the depth of maximum response to temperature variations as computed numerically with STiC. The response function to temperature was calculated for every inverted pixel. The distributions of the optical depth of peak response to temperature of the $h_{2r}$ and $k_{2r}$ features are shown in the third panel of Fig.~\ref{Figure:7} for observations 7 and 9. For both observations there are clearly two groups of pixels; one for which the formation depth of the $h_{2r}$ and $k_{2r}$ features is in the upper atmosphere and another for which it is in the middle chromosphere. This was an expected result, as there is the possibility that the emission peaks of the \ion{Mg}{II} h \& k lines form deeper in the atmosphere \citep{2013ApJ...772...90L}. Pixels pertaining to the second aforementioned group were discarded as they formed deeper in the atmosphere, arguably below the height of the coronal rain material, with the division between both groups set at $\log{\tau_{500}}=-4.5$, where the distribution of formation depths of observation 9 has a maximum (see bottom right panel of Fig.~\ref{Figure:7}). The coronal rain material in observation 7 is much more abundant than the small rain event of observation 9; therefore, the coronal rain pixels in observation 7 mostly belong  to the first group mentioned above, with the opposite being true for observation 9. With the very small amount of pixels in observation 9 forming at upper layers of the atmosphere, this observation was not used. In observation 7, the discarded pixels belong to the edges of the coronal rain material (yellow-coloured pixels in the upper middle panel of Fig.~\ref{Figure:7}), where (arguably) the coronal rain material exhibits a lower opacity; therefore, the formation height of the $h_{2r}$ and $k_{2r}$ features is in the middle chromosphere, not in the coronal rain itself.

\subsection{Velocities}

\begin{figure*}[h!]
 \centering
 \includegraphics[width=18cm]{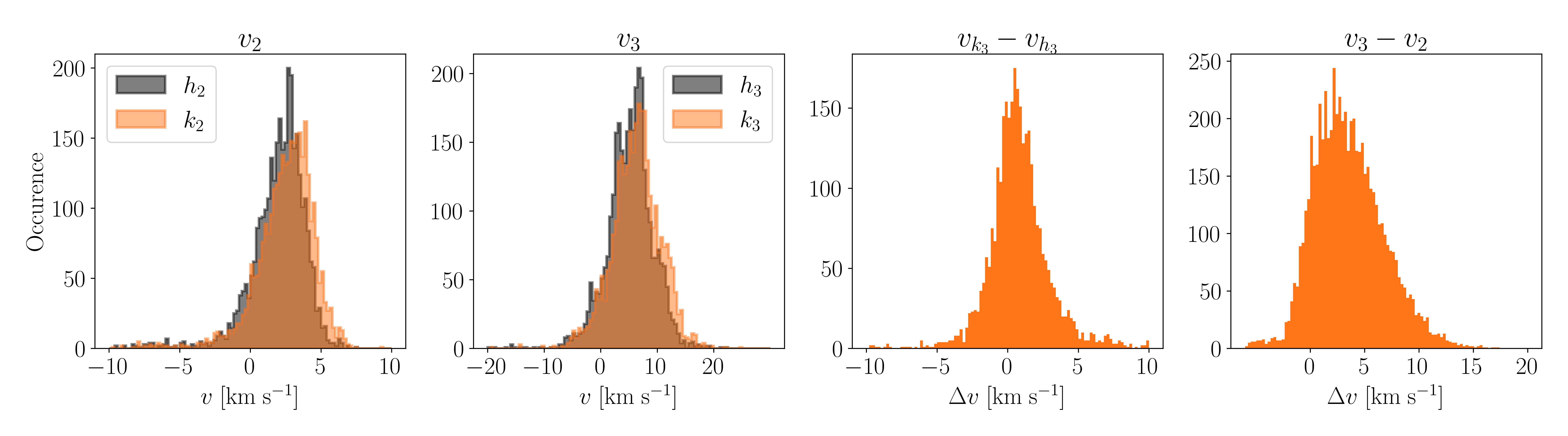}
 \includegraphics[width=18cm]{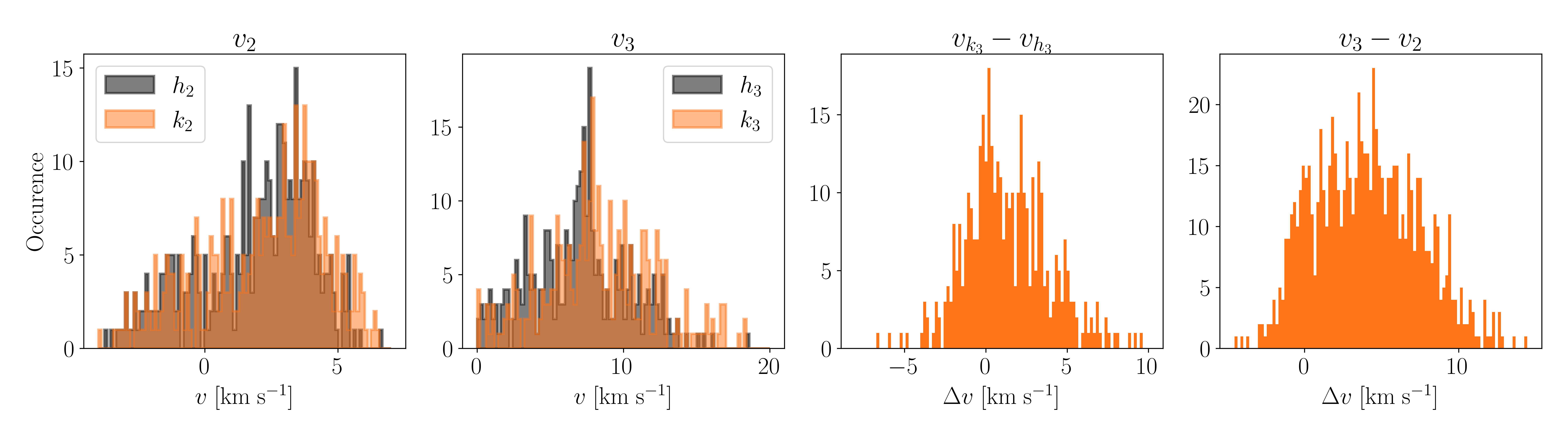}
 \includegraphics[width=18cm]{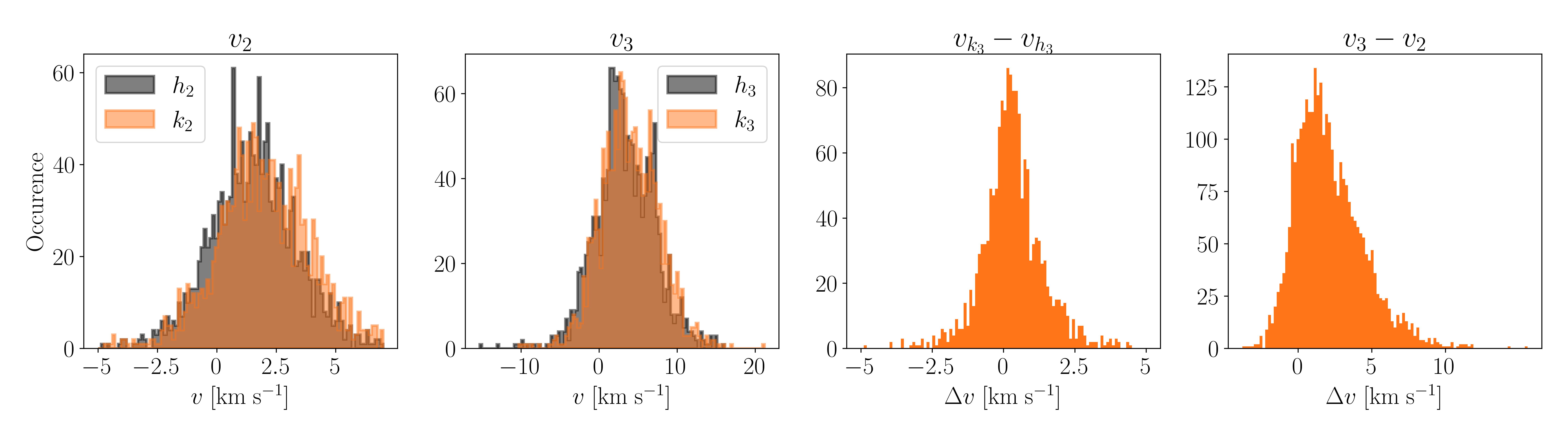}
 \includegraphics[width=18cm]{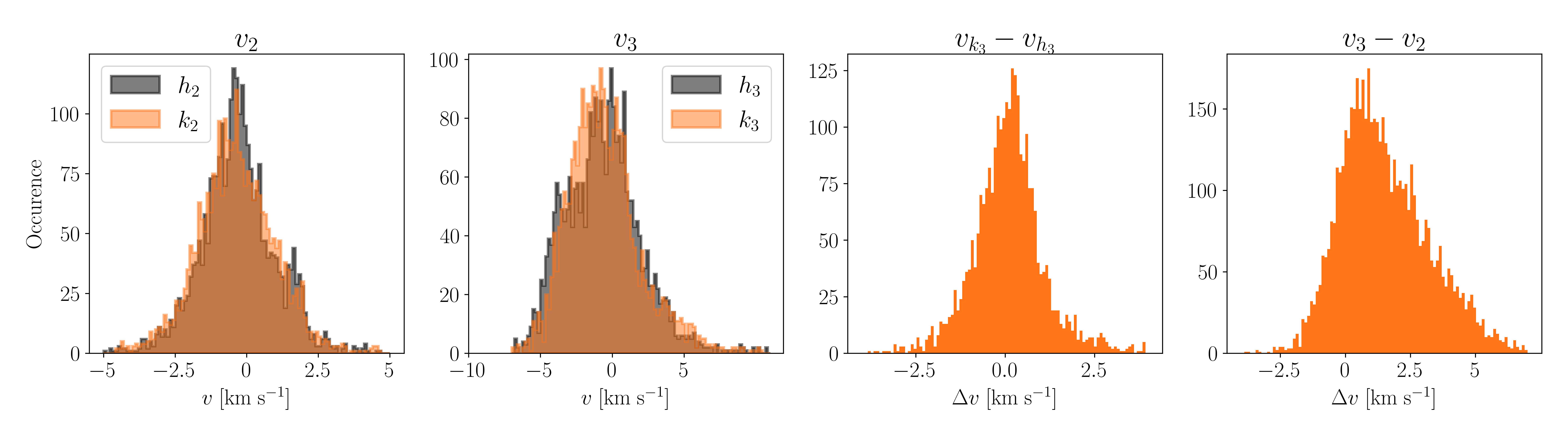}

 \caption{Velocity diagnostics. Each row represents a different observation, from observation 1 to 4 in ascending order from the top to the bottom. The first two columns represent the distribution of the LoS velocity as measured in the position of the emission peaks (first column) and in the line core (second column). The third column shows the distribution of the relative values of the line core velocities between the h \& k lines, and the fourth column shows the relative velocity as measured in the line core and the emission peaks.  }
 \label{Figure:5}%
\end{figure*}

\begin{figure*}[h!]
 \centering
 \includegraphics[width=18cm]{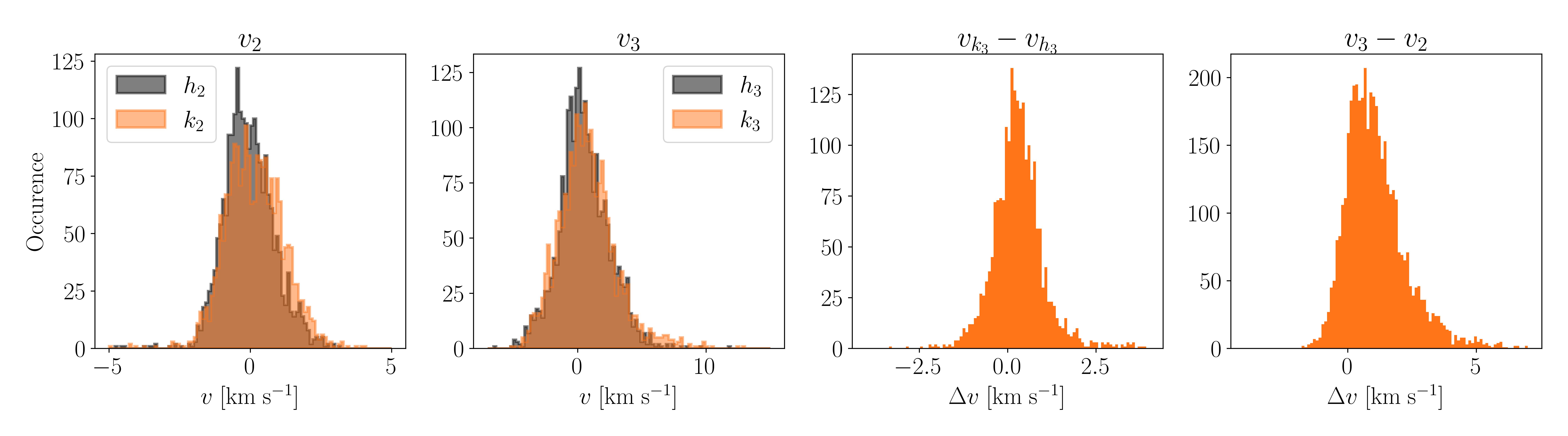}
 \includegraphics[width=18cm]{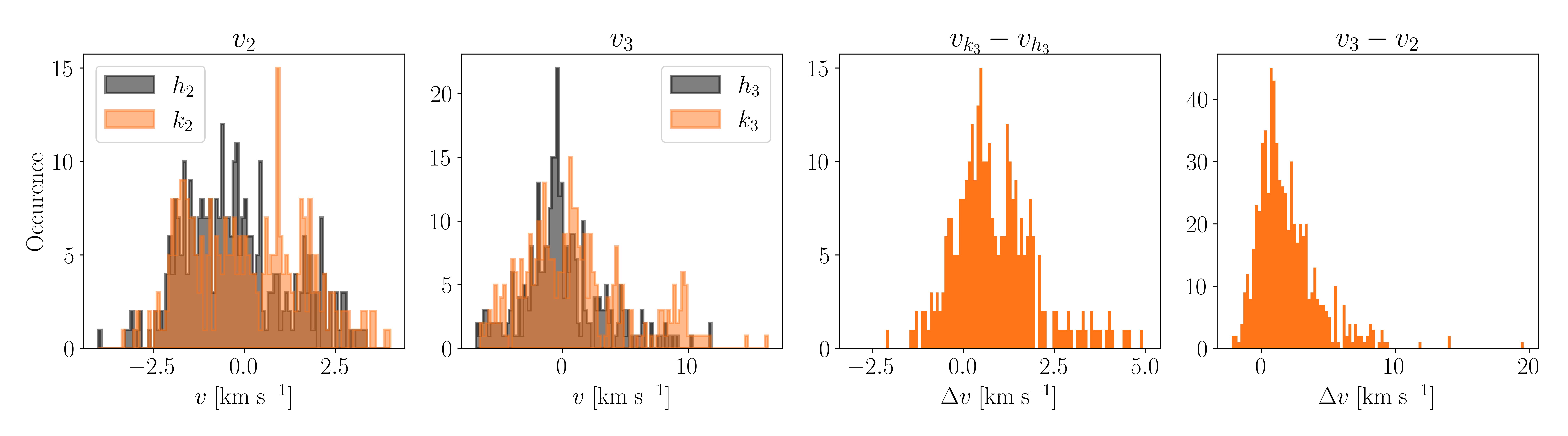}
 \includegraphics[width=18cm]{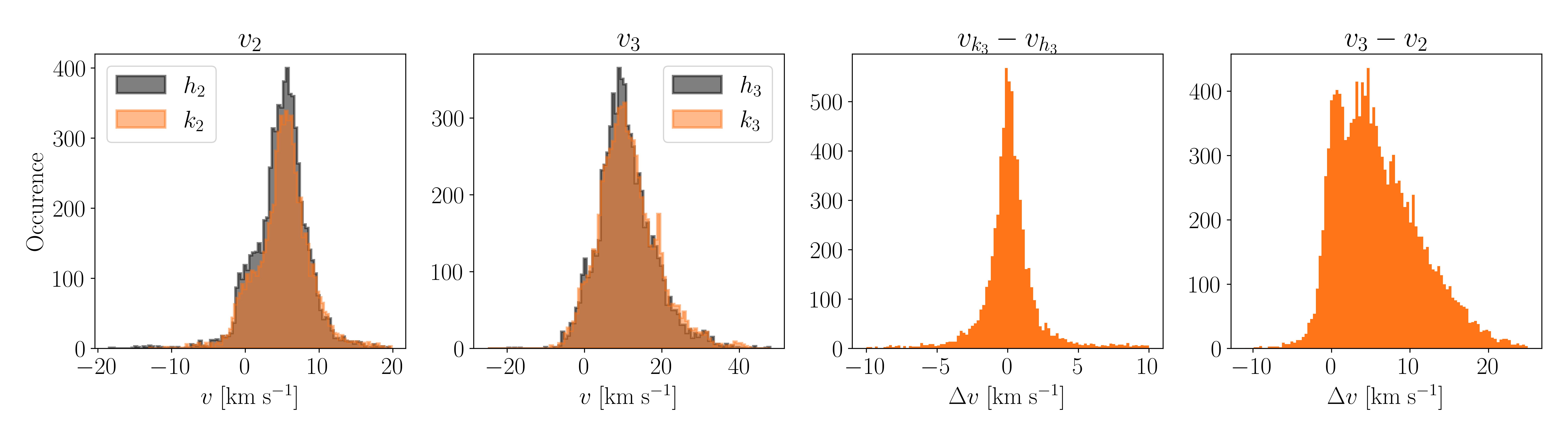}
 \includegraphics[width=18cm]{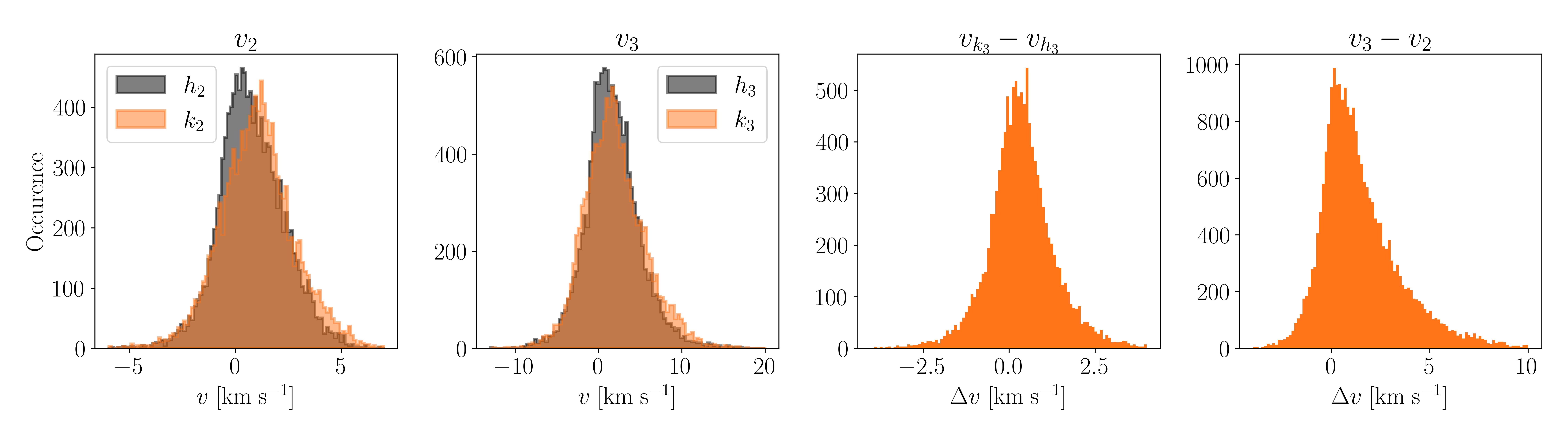}

 \caption{Same as Fig.~\ref{Figure:5}, for observations 5 to 8. }
 \label{Figure:6}%
\end{figure*}

The methods described in Sects.~\ref{sect:43} and~\ref{sect:44} were applied to all the observations, with the results displayed in Figs.~\ref{Figure:5} and~\ref{Figure:6}. Positive values of the velocity in the first two panels correspond to movements away from the observer. In the first two panels of each row, the velocities at the formation height of the emission peaks (first panels) and of the line centre (second panels) are mostly positive, indicating that the material is falling towards the chromosphere. The derived velocity distributions for observations 4, 5, and 6 do have a sizeable negative proportion of values. As they are all close together in time and location (see Table~\ref{table:Table1}), this could be caused by the orientation of the coronal loops respect to the observer. The measured velocities with respect to the observer are rather small in comparison to the values reported by studies in the \ion{Ca}{II} H line and in the H$\alpha$ line \citep{2010ApJ...716..154A,2012ApJ...745..152A,2015ApJ...806...81A,2020A&A...633A..11F,2021A&A...650A..71K}. The reason for this discrepancy is that in all observations reported here the motion of the coronal rain clump is constrained mostly to the plane of the sky in a perpendicular direction to the observations.

The larger opacity of the \ion{Mg}{II} k line leads to a slightly higher formation height than the h line \citep{2013ApJ...772...90L}. Additionally, the emission peaks of each line form deeper than the line cores. Assuming that this is also true in the case of coronal rain, it is (in principle) possible to estimate velocity differences between different parts of coronal rain clumps. Those are shown in the third and fourth columns of Figs.~\ref{Figure:5} and~\ref{Figure:6}. The relative velocities between different heights are measured with the expression:

\begin{equation} \label{eq:1}
V_r = |v_h|-|v_l|\times \mathrm{sign}(v_h)\times\mathrm{sign}(v_l), \end{equation} 

\noindent where $v_h$ is the velocity assumed to be measured higher in the atmosphere and $v_l$ is the velocity assumed to be measured at a lower height. With Eq.~(\ref{eq:1}), a positive relative velocity means that the higher layer is moving faster towards or away from the observer than the velocity at the lower layer and a negative velocity means the opposite is true.

The relative velocity between the core of the h \& k lines is relatively small, as the cores of the lines tend to form very close to each other \citep{2013ApJ...772...90L}. However, the distributions seem to all be slightly shifted towards positive values, indicating that the plasma at the outer layers of the coronal rain clump is moving faster than the plasma below it. The right-most panels of Figs.~\ref{Figure:5} and~\ref{Figure:6} show the relative velocity between the core of each line and its emission peaks, again estimated using Eq.~(\ref{eq:1}). The relative velocities are slightly larger than the ones measured between the two line cores, with the shift towards positive values is even larger than that between the line cores, indicating that the layer where the line core is formed is moving with larger speeds than the layer where the emission peaks are formed. Such result should be confirmed and justified in the future thanks to further observations and coronal rain simulations.

\subsection{Temperatures and densities}

\begin{figure*}[h!]
 \centering
 \includegraphics[width=4.5cm]{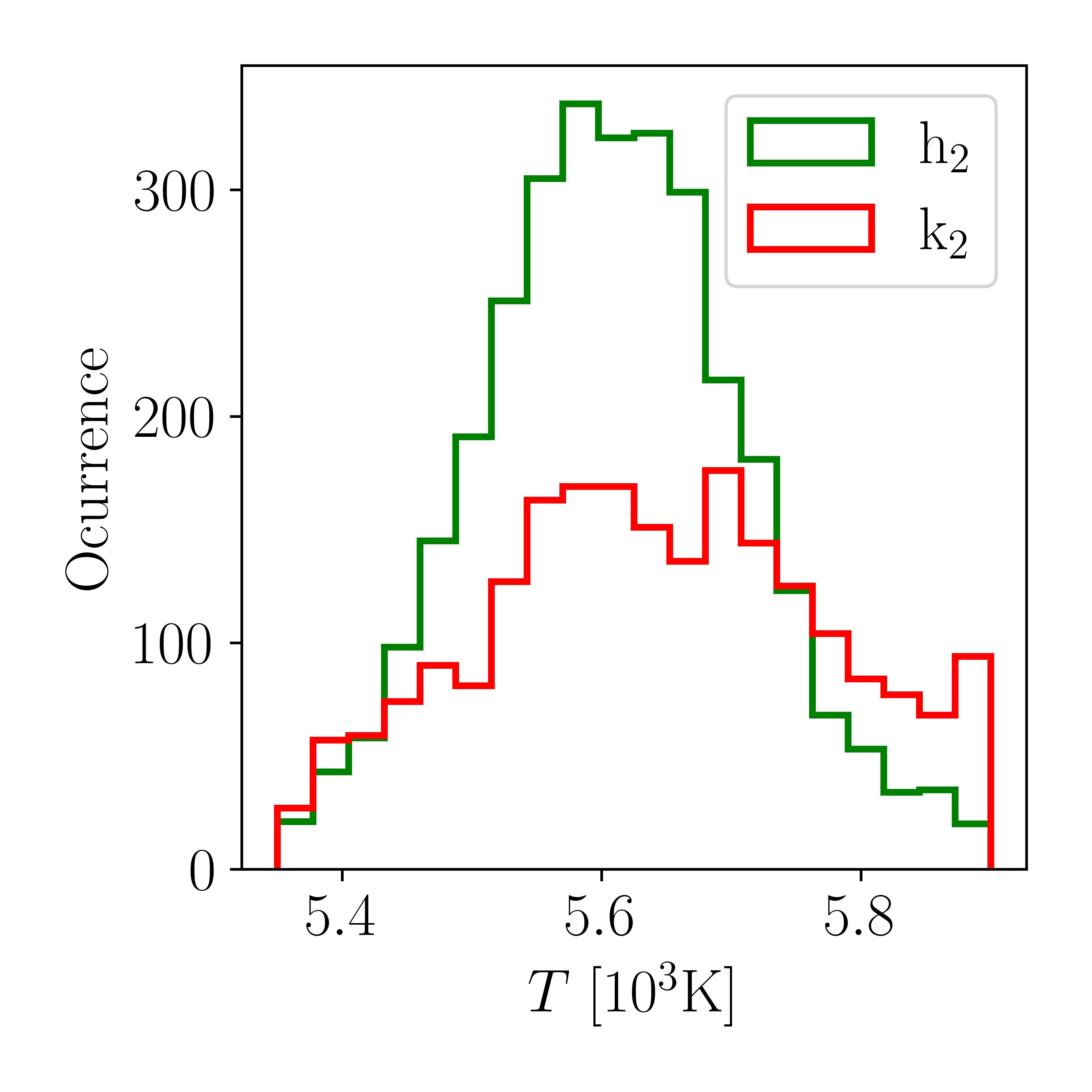}
 \includegraphics[width=4.5cm]{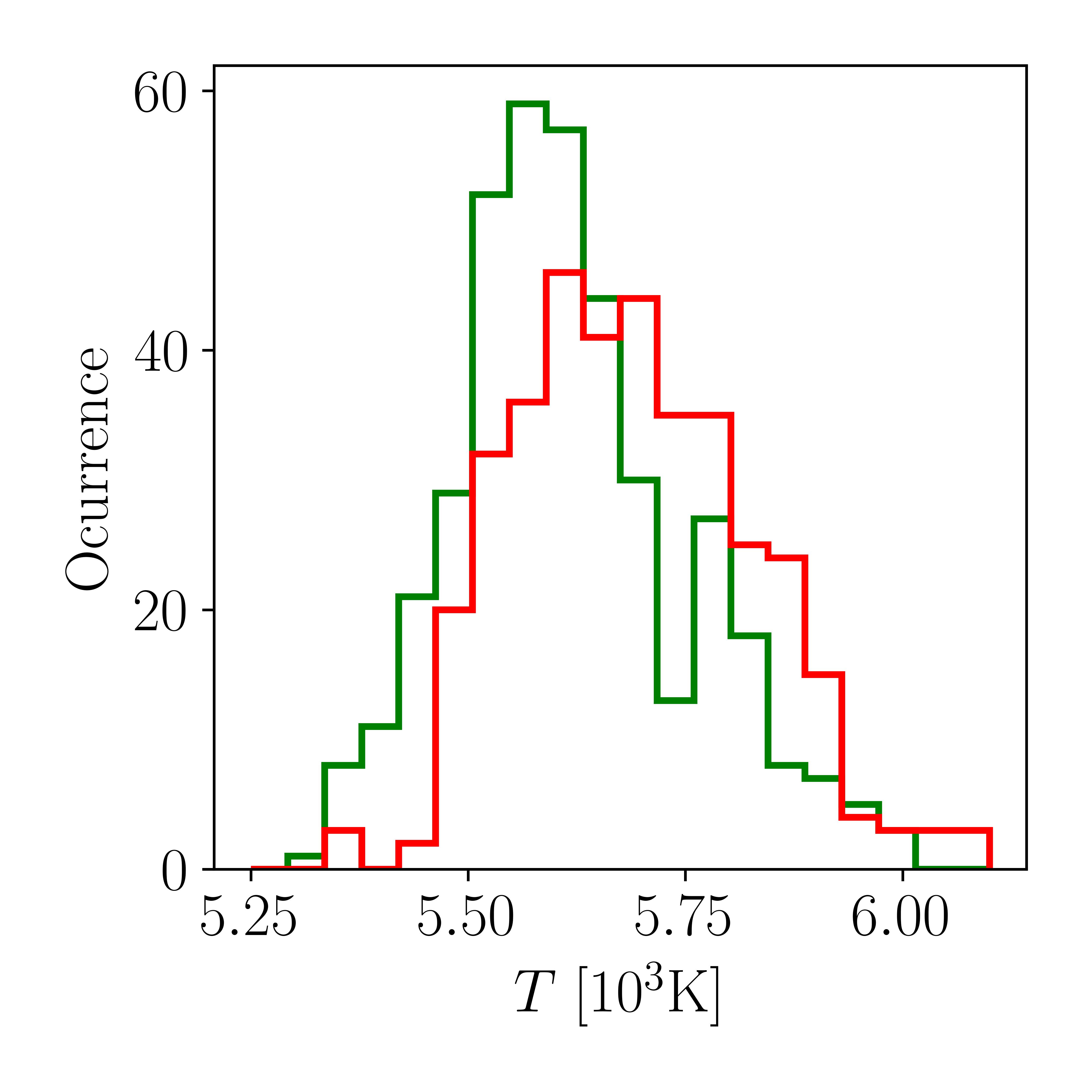}
 \includegraphics[width=4.5cm]{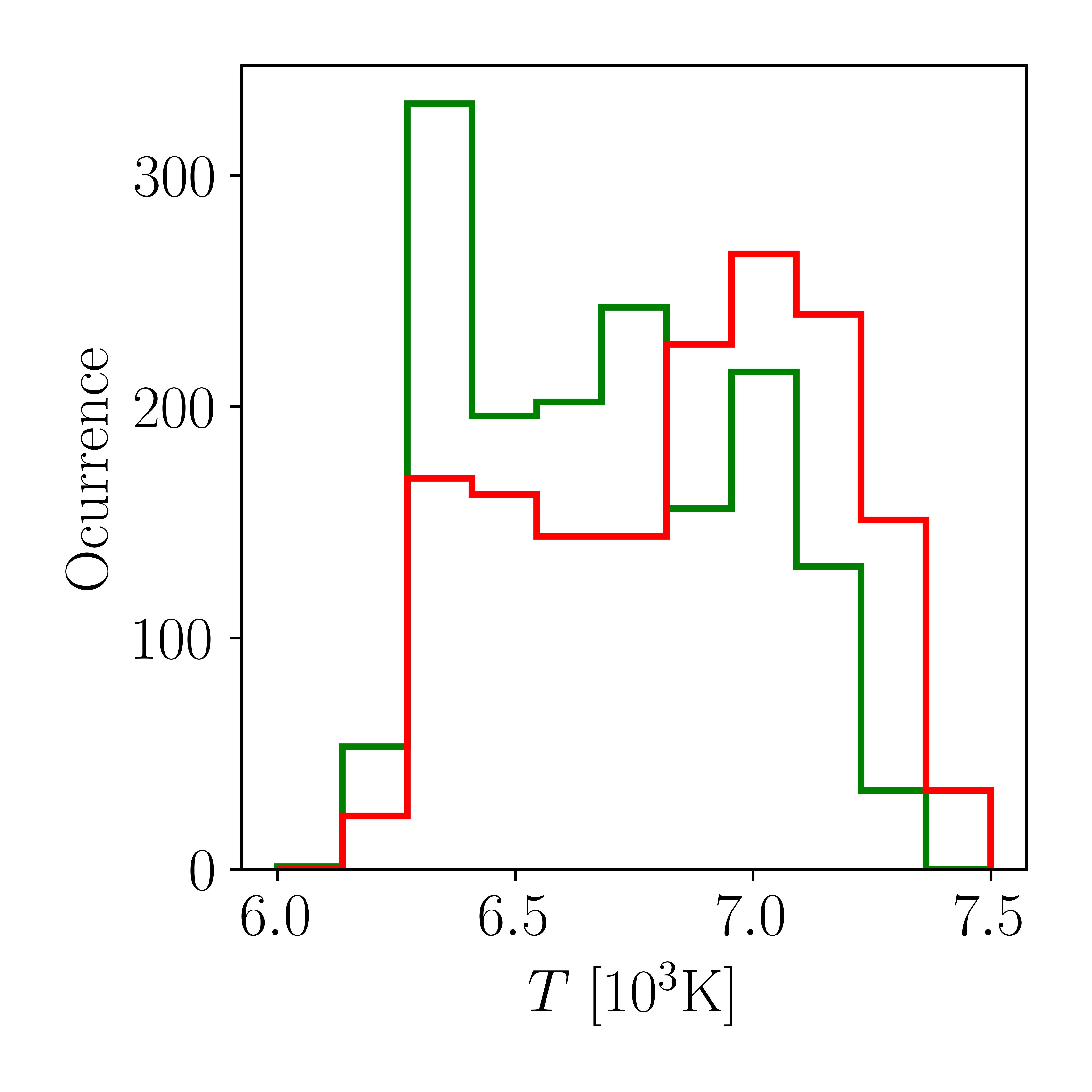}
 \includegraphics[width=4.5cm]{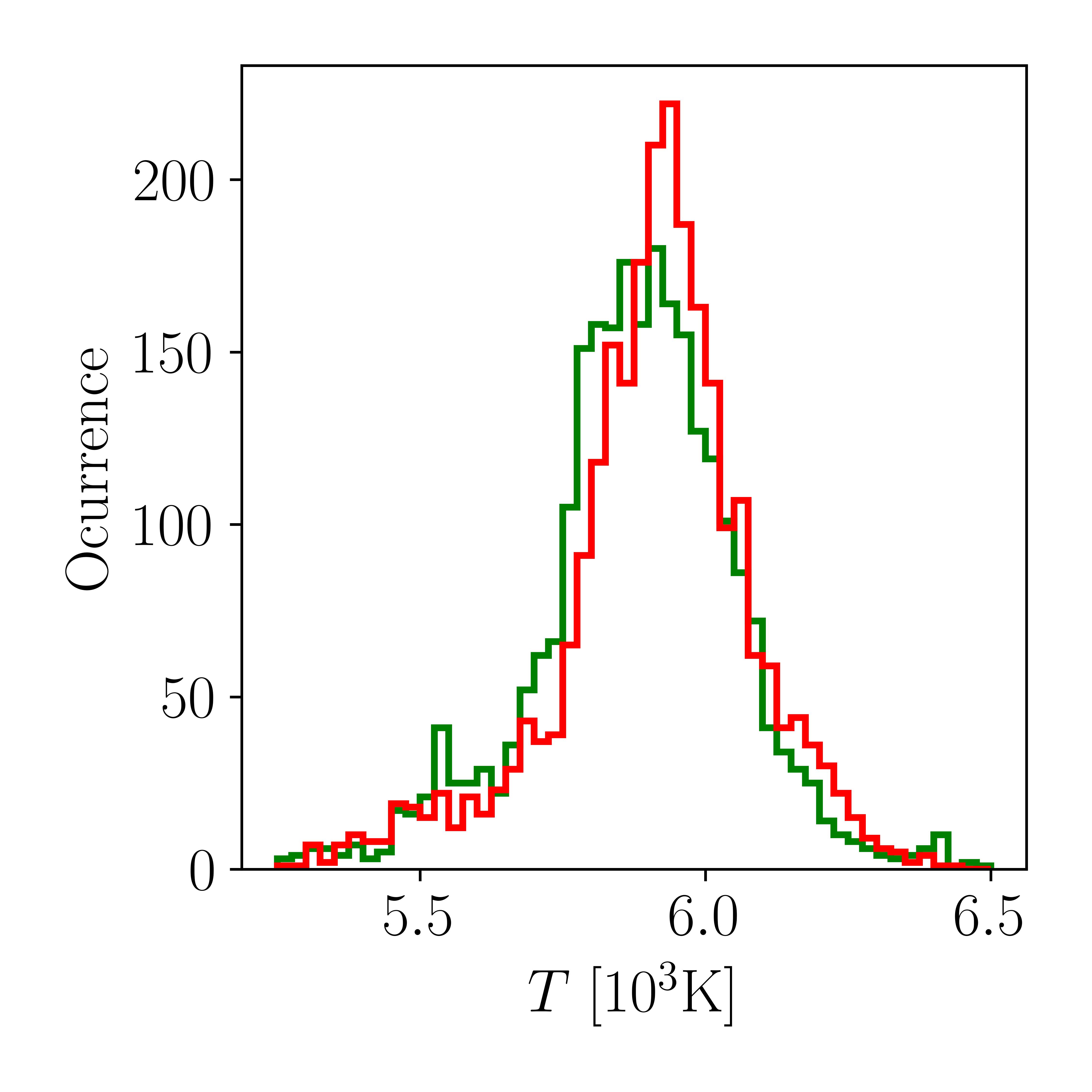}
 \includegraphics[width=4.5cm]{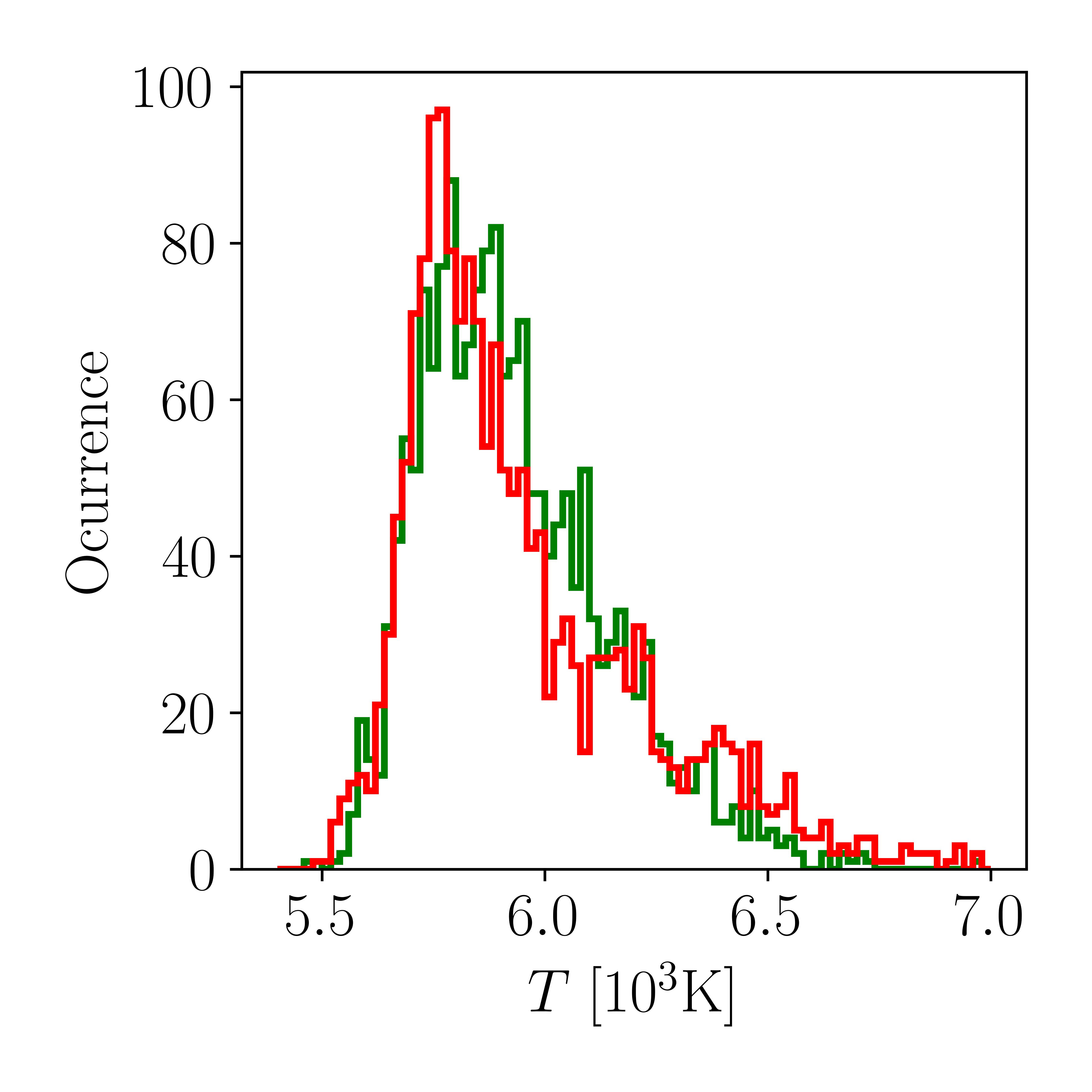}
 \includegraphics[width=4.5cm]{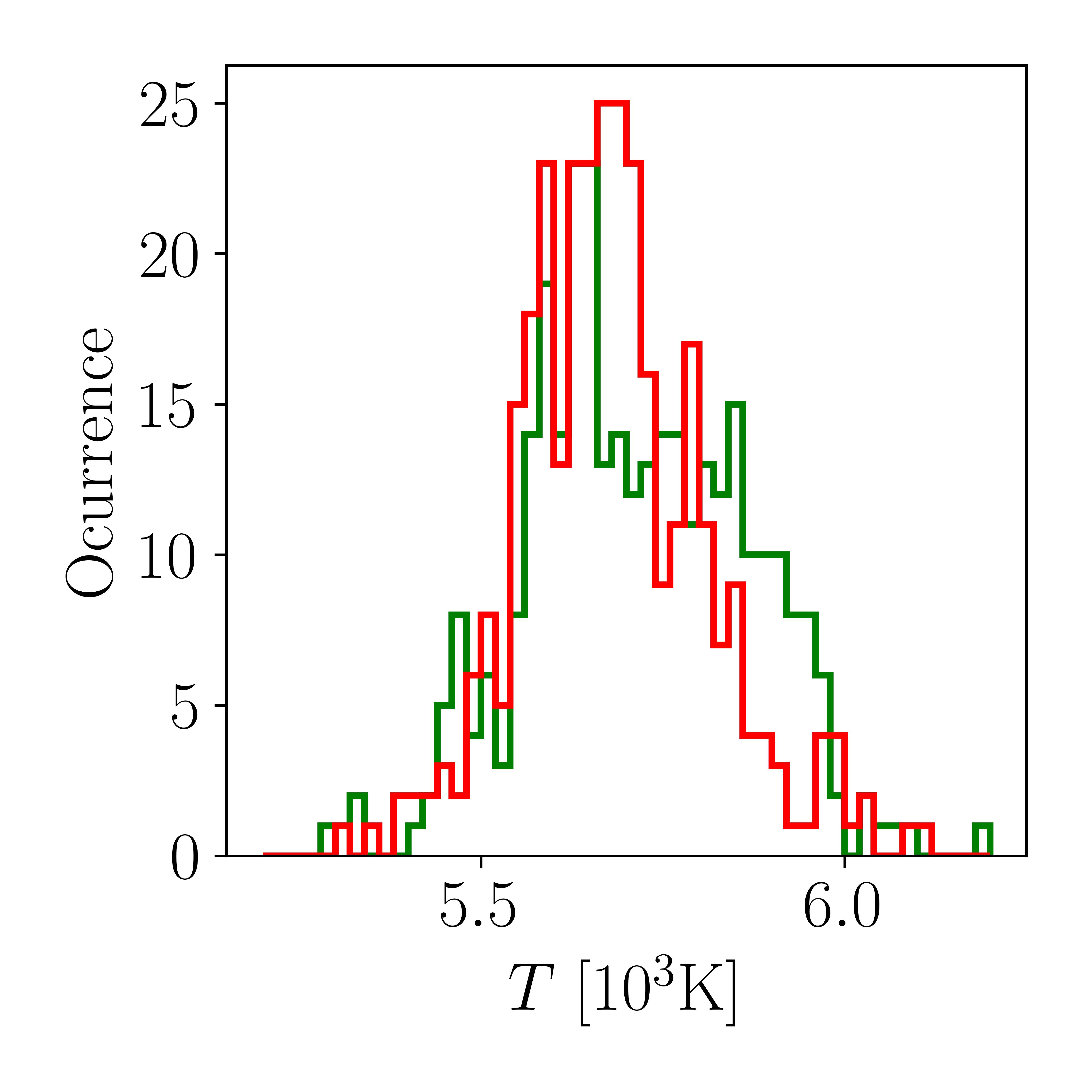}
 \includegraphics[width=4.5cm]{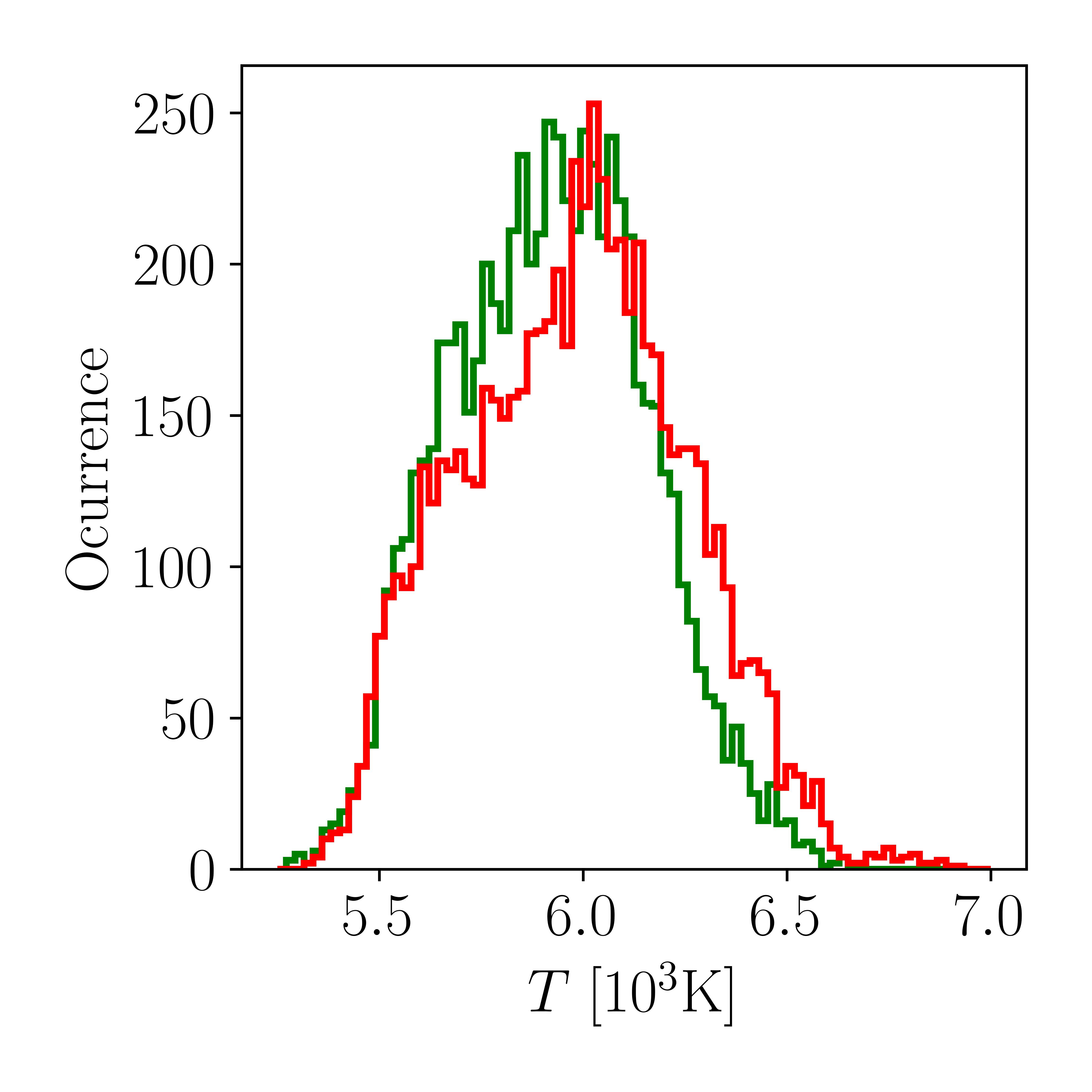}
 \includegraphics[width=4.5cm]{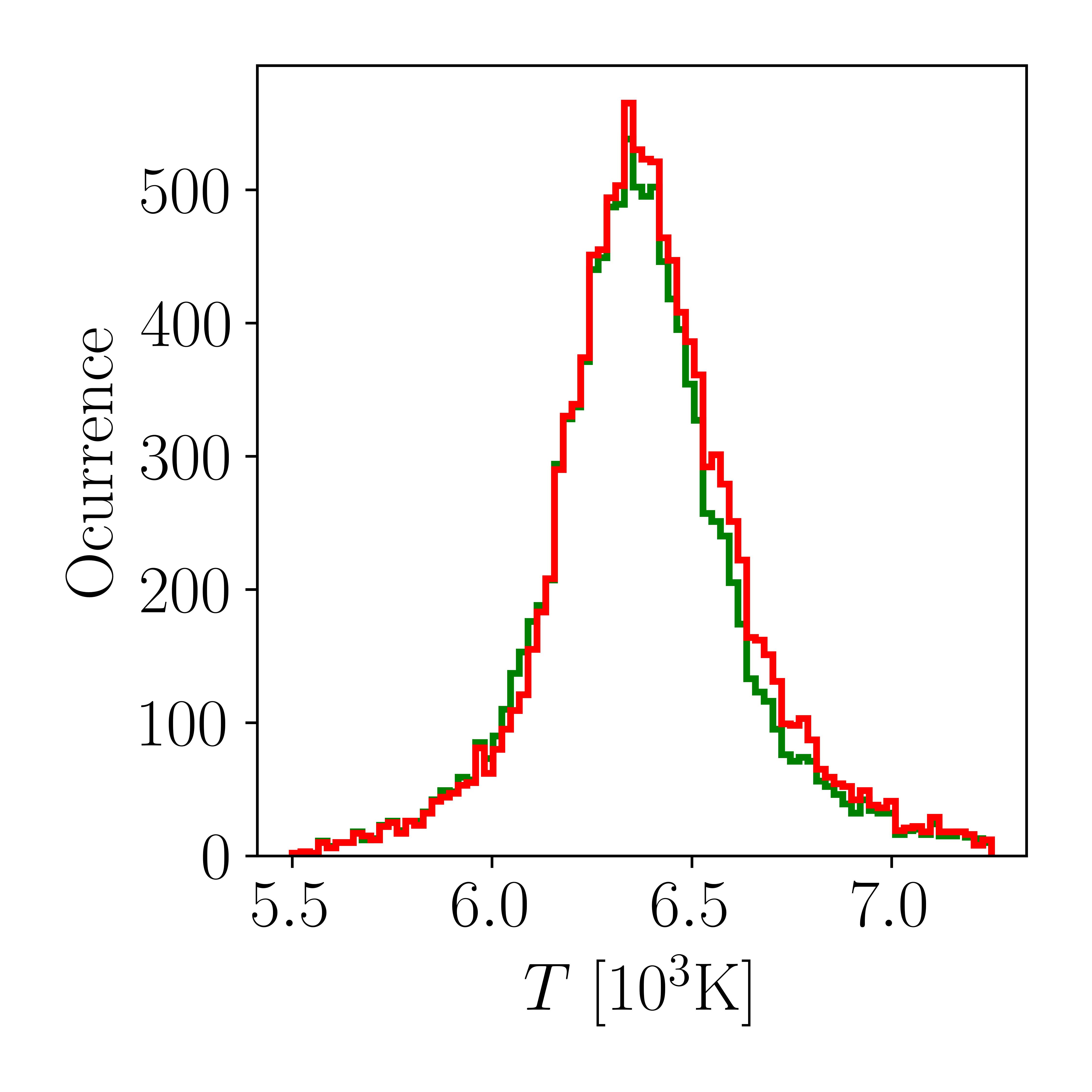}

 \caption{Temperature diagnostics. From left to right and top to bottom: Histogram of the average temperature at the depth of maximum response to temperature of the $h_2$ and $k_2$ features for observations 1 through 8. }
 \label{Figure:8}%
\end{figure*}

The temperature inversions were performed with the methodology described in Sect.~\ref{sec:Data}. For each pixel in the different observations inversions (for which both the $h_2$ and $k_2$ are formed in the upper atmosphere), Fig.~\ref{Figure:8} shows the temperature distribution at the height of formation of these features for each of the observations. Observation 9 was discarded because for its pixels, both the $h_2$ and $k_2$ features are mostly formed in the middle chromosphere.

The inverted temperatures are consistent between the different observations and all remain within the narrow temperature range between 5200 K up to 7500 K. This could indicate that in the different condensation events, the rain becomes visible in the \ion{Mg}{II} h \& k lines in a very specific moment during its cooling phase and that different condensation events pertaining to independent coronal loops reach similar thermal conditions. The inverted temperature values are consistent with the temperature at the denser core of coronal rain found in simulations \citep{2022ApJ...926L..29A}, which could mean that that is where the emission peaks of the \ion{Mg}{II} h \& k lines are being formed. The inverted temperature of the $k_2$ feature seems to be slightly larger than that inferred at the $h_2$ feature (with the exception of observation 6), indicating that the $h_2$ forms closer to the core of the coronal rain clump. Some of the temperature distributions show apparent peaks centred at multiple temperature values. For example, the temperature distribution at the formation height of the $k_2$ feature for observation 3 has a peak around 6500~K and another apparent peak at 7100~K. The temperature distributions for observation 7 have multiple peaks for both the $h_2$ and $k_2$ features. This multiplicity in distribution peaks can be caused by the variation of the geometrical height of formation of the spectral features even for a single observation, where they might have their formation at locations of different temperatures. Additionally, observation 7 stems from a flare event which led to the formation of at least three condensation events that crossed the position of the slit. Small differences in the properties of the plasma constituting these events could also serve as another cause for the variety of peaks in the corresponding temperature distribution.
 
\begin{figure*}[h!]
 \centering
 \includegraphics[width=4.5cm]{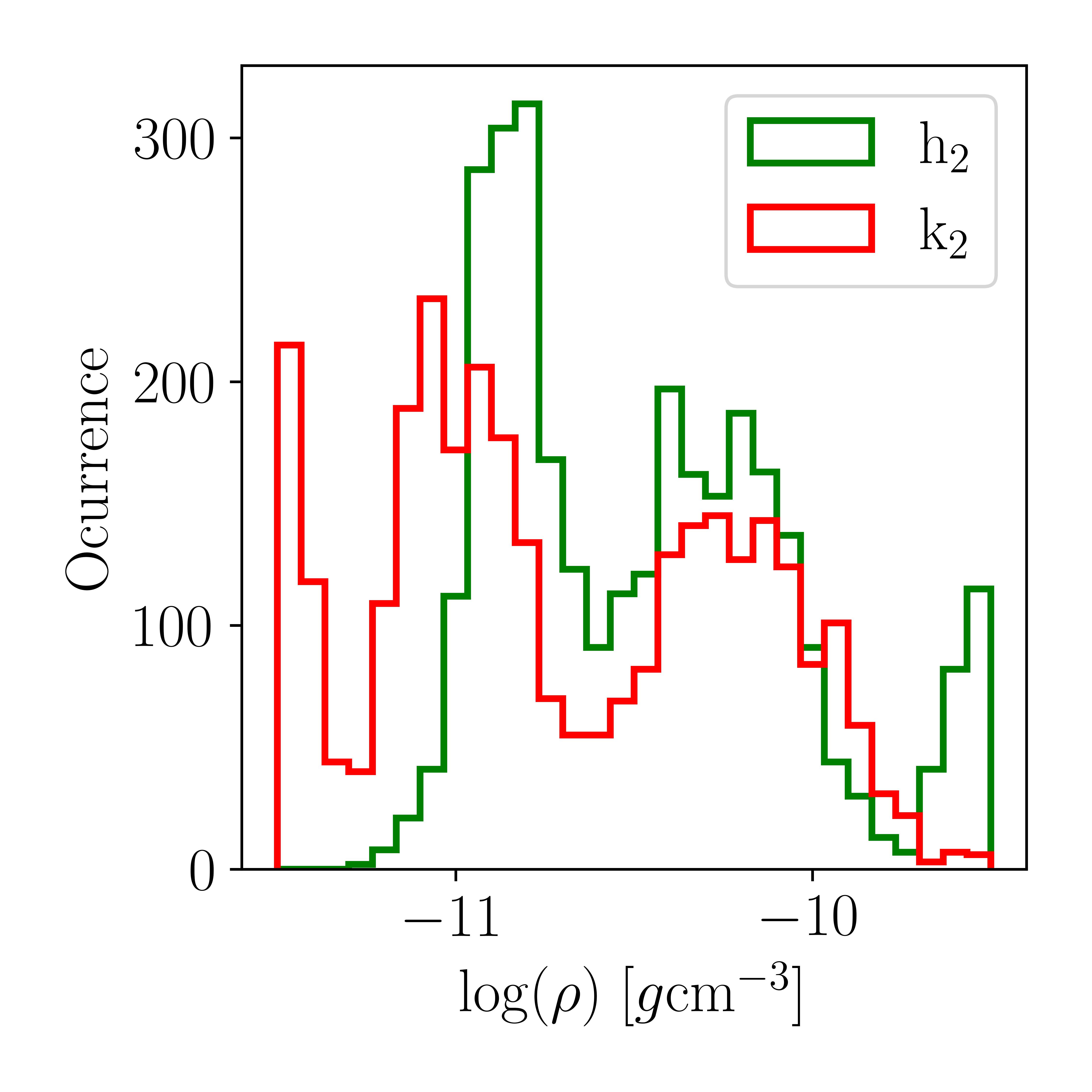}
 \includegraphics[width=4.5cm]{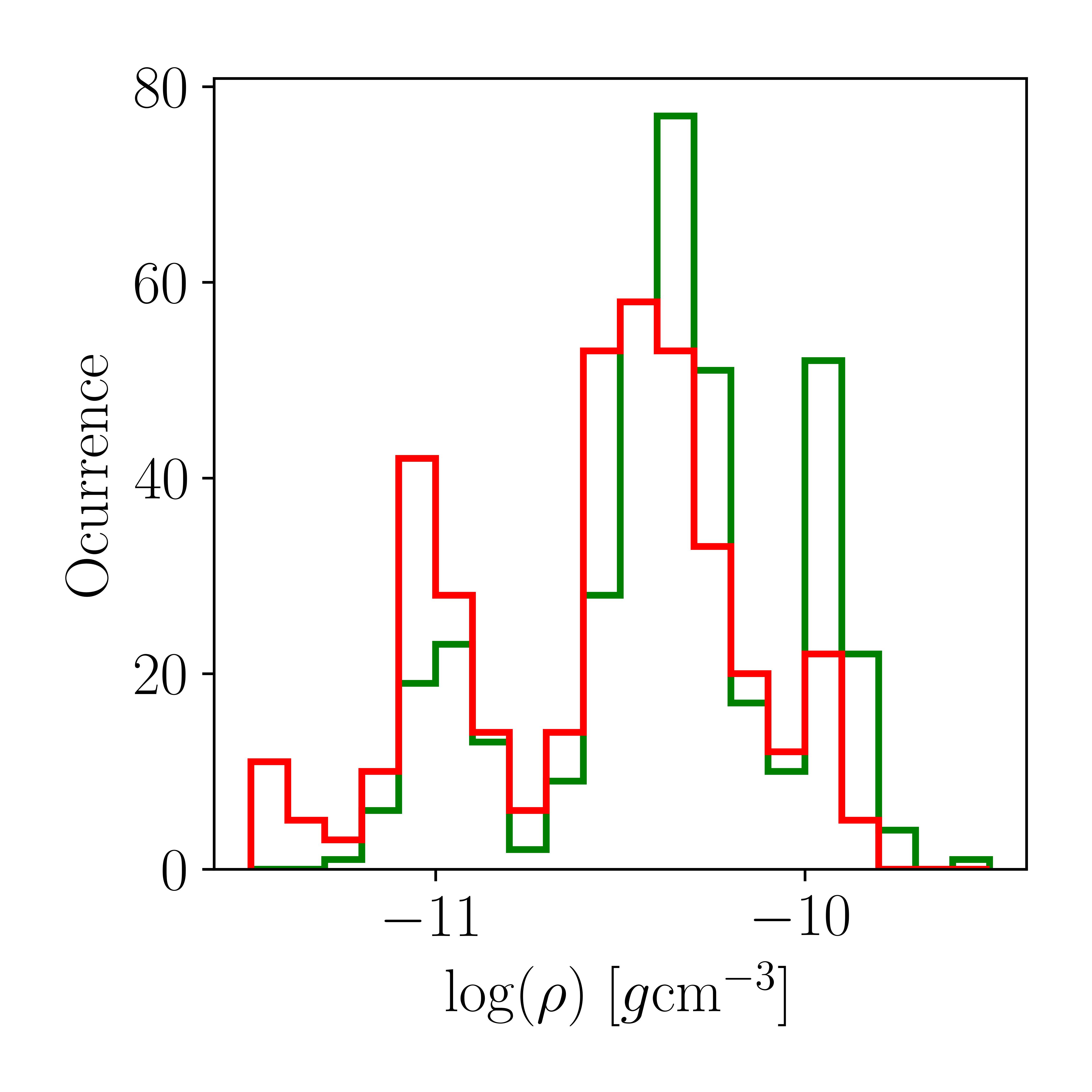}
 \includegraphics[width=4.5cm]{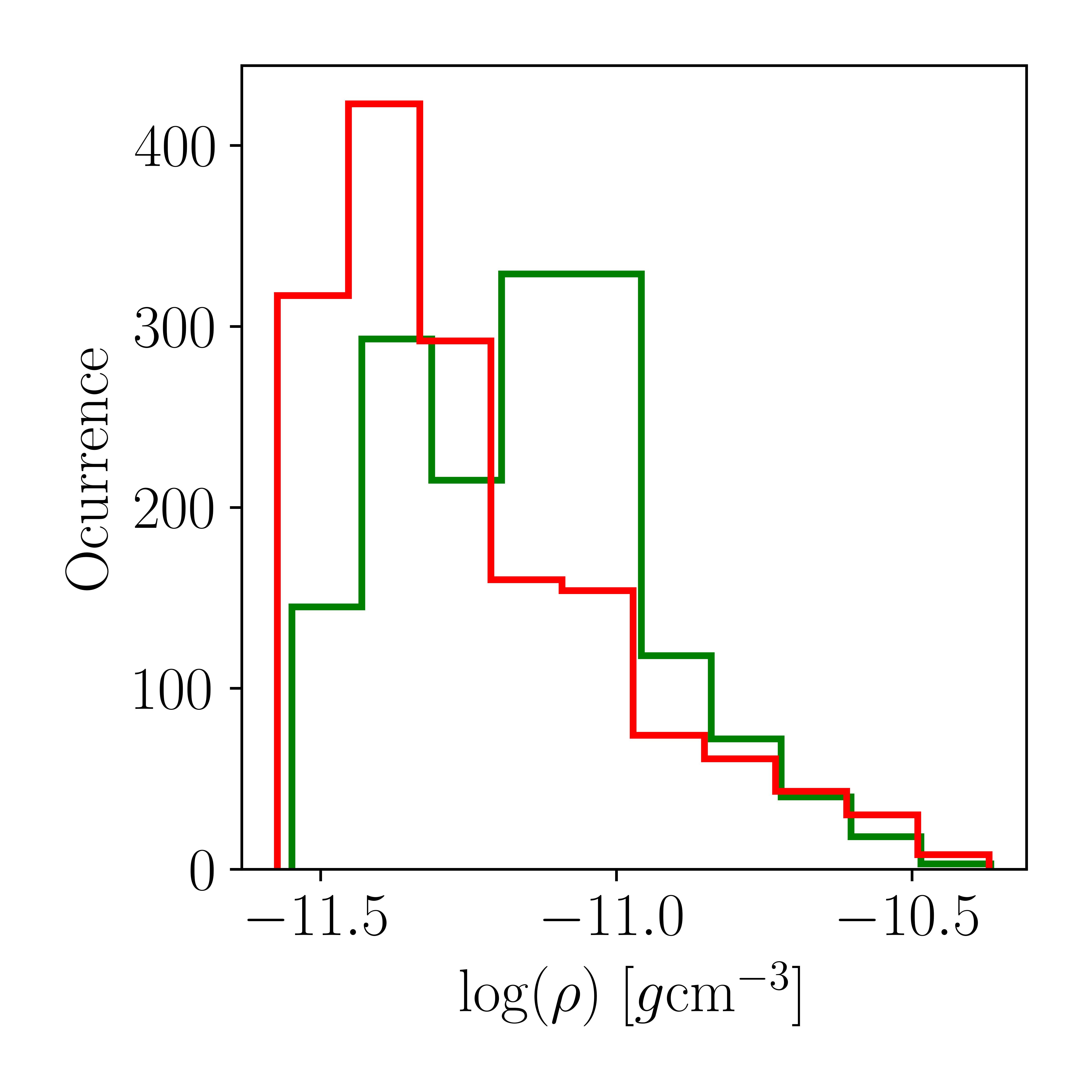}
 \includegraphics[width=4.5cm]{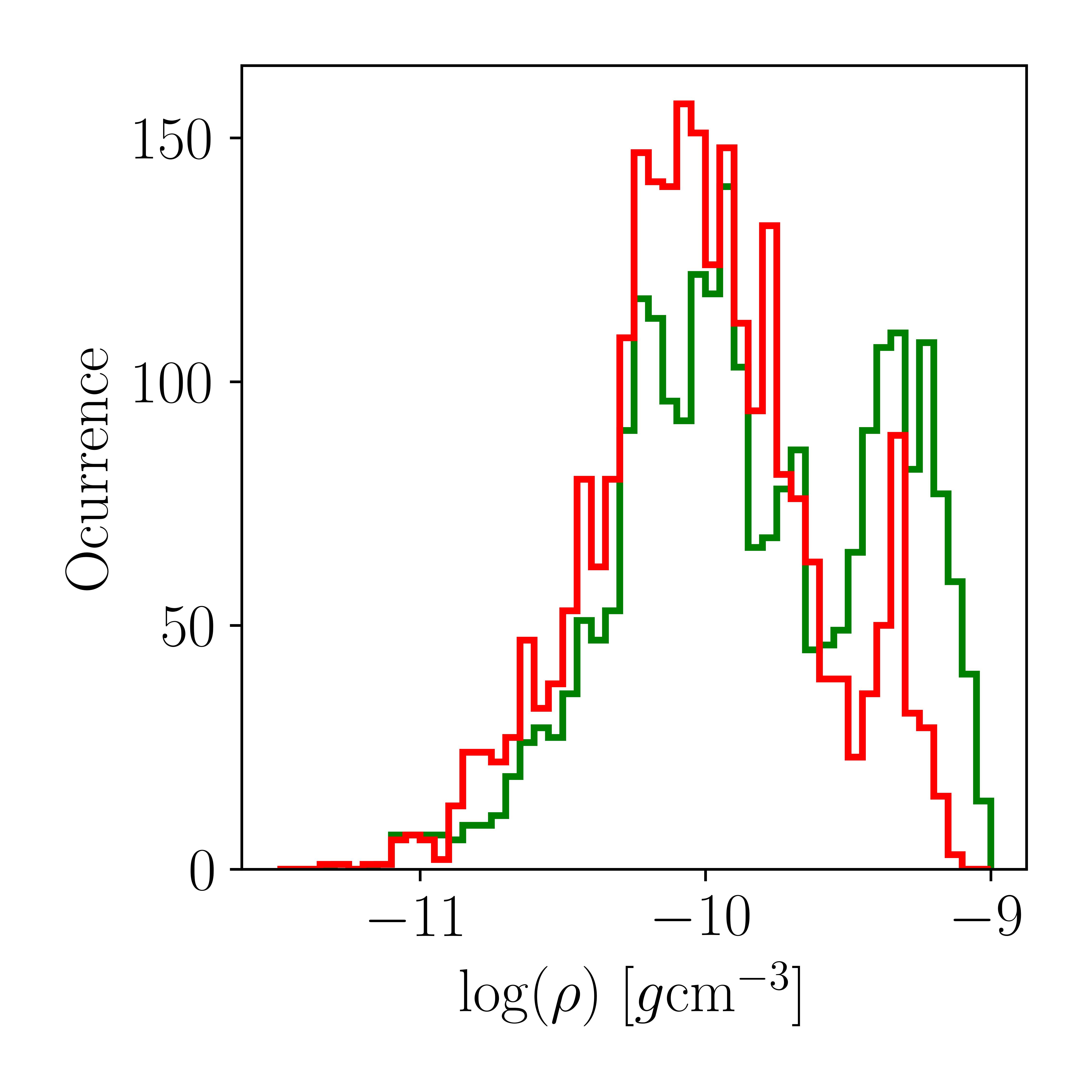}
 \includegraphics[width=4.5cm]{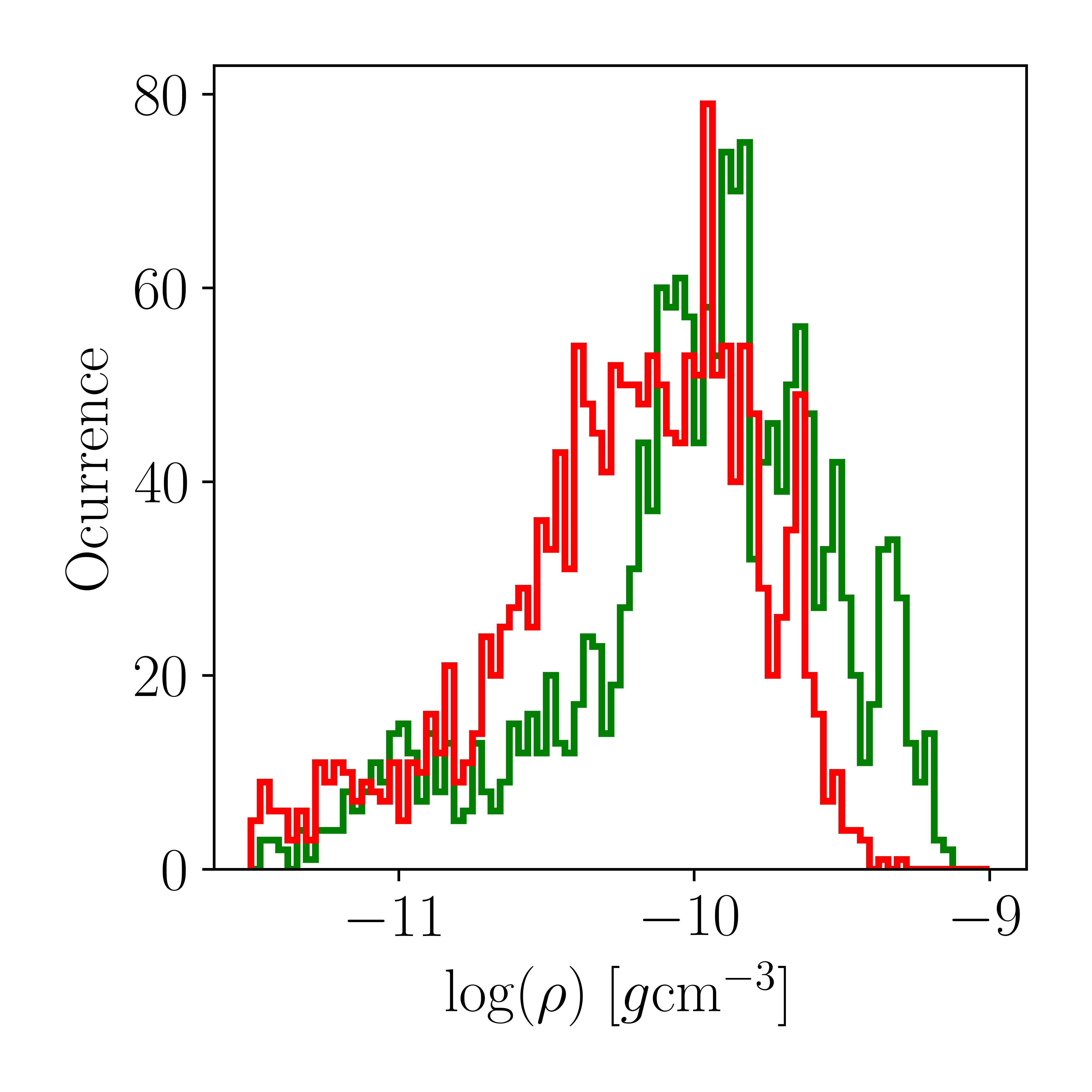}
 \includegraphics[width=4.5cm]{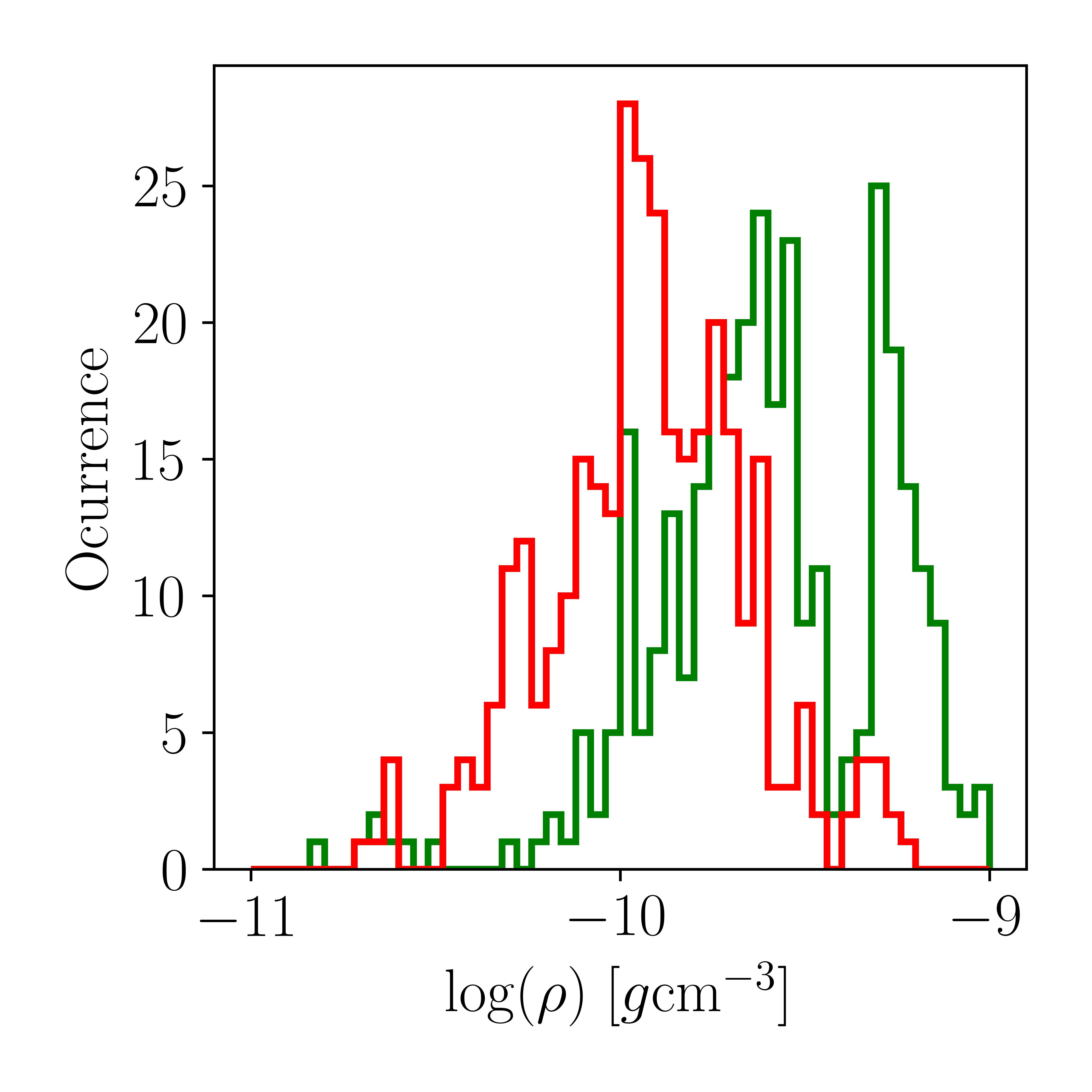}
 \includegraphics[width=4.5cm]{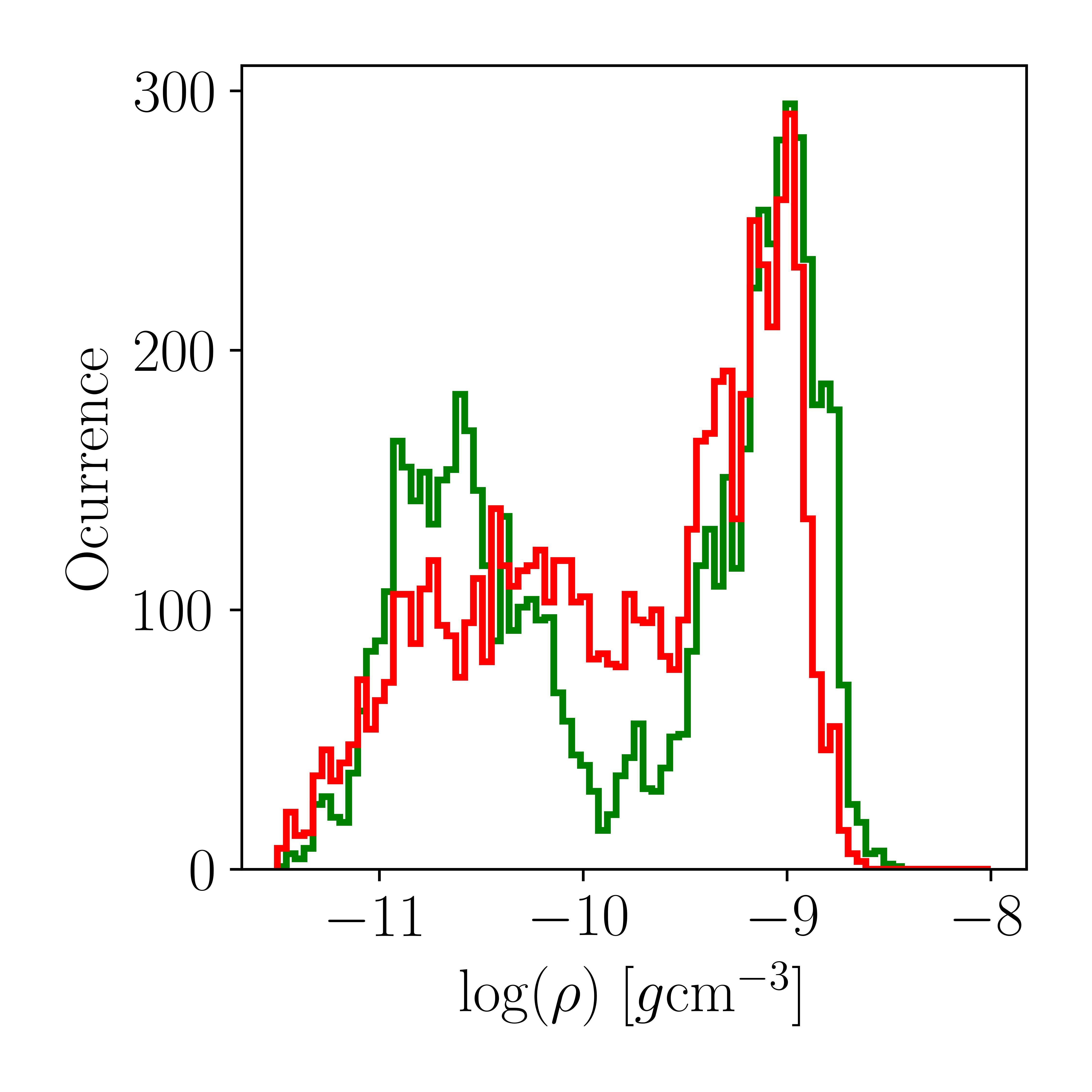}
 \includegraphics[width=4.5cm]{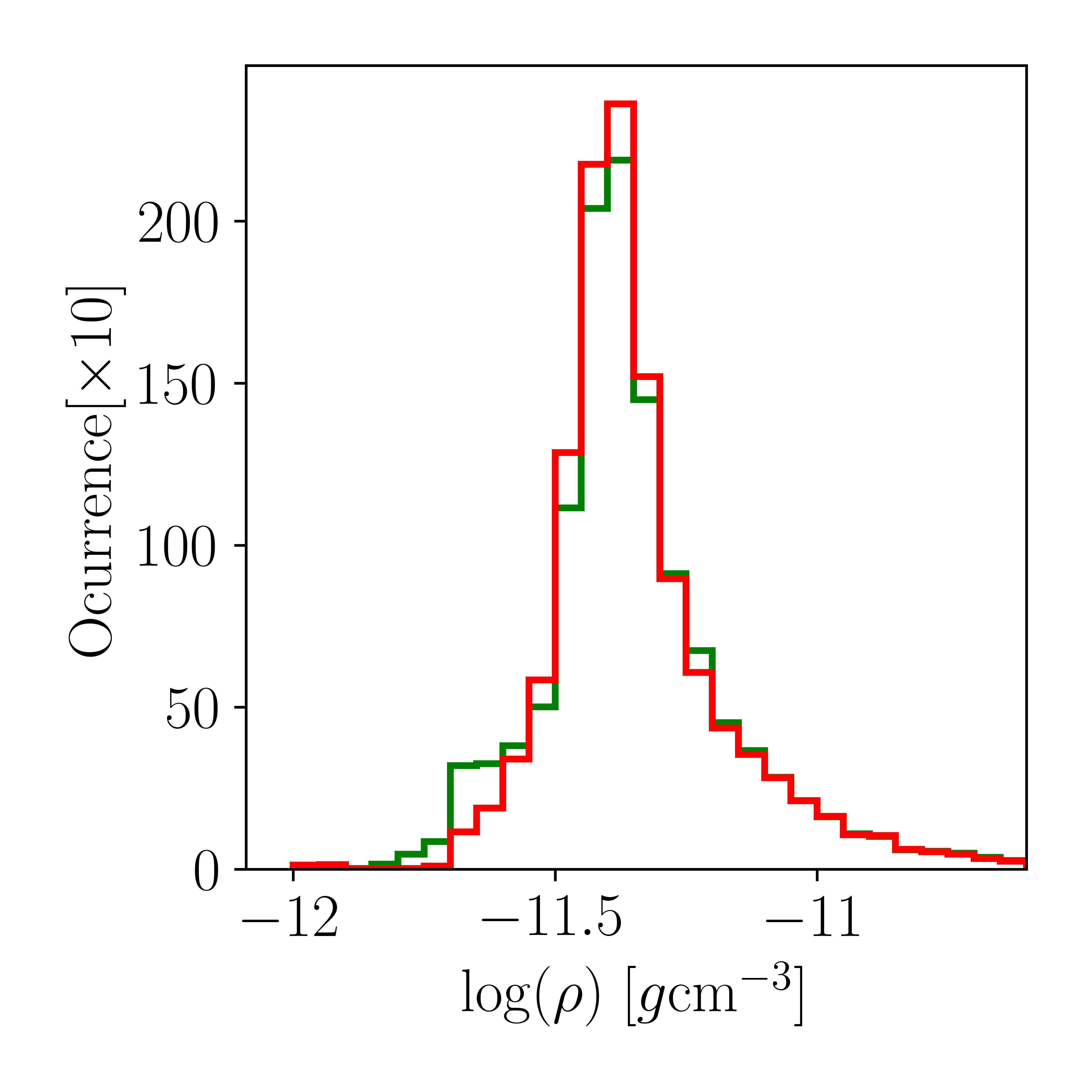}

 \caption{Density diagnostics. Same as Fig.~\ref{Figure:8}, but for densities. }
 \label{Figure:9}%
\end{figure*}

Figure~\ref{Figure:9} shows the derived densities, which are also similar to those at the denser cores of the coronal rain plasma found in simulations \citep{2022ApJ...926L..29A} and with values obtained from estimates of the density based on the correlation between the emission measure and intensity in the H$\alpha$ line \citep{2020A&A...633A..11F}. Similarly to the temperature values, there seems to be some shift between the densities inferred in the $h_2$ and $k_2$ features, with denser values at the formation height of the $h_2$ feature being inferred.

\section{Conclusions} \label{sec:conc}

 The strongly stratified heating that coronal loops are subject to can give rise to the ubiquitously-observed phenomenon of coronal rain. The multi-thermal nature of coronal rain allows for it to be studied with lines formed in the chromosphere, the transition region and the corona. This study has made use of successful observations of coronal rain formation in the \ion{Mg}{II} h \& k lines as observed with the spectrograph mounted in the IRIS satellite.

 As opposed to the off-limb observations, which rely in the inference of the temperature assigning a single temperature value to the emitting plasma, inversions can in theory provide a depth-dependent temperature stratification. The non-LTE code STiC was used for the inversion of coronal rain spectra in nine different active region observations obtained on the solar disc. The sensitivity of the \ion{Mg}{II} h \& k lines allowed for the inference of the temperature of coronal rain clumps using this alternative method. The results largely agree with inferences made using thermal width estimates from the emission line profiles. The results also agree with those obtained in numerical simulations of coronal loops undergoing catastrophic cooling. The results are specially aligned with the cool, dense cores of such studies, which might suggest that this is where the \ion{Mg}{II} h \& k lines are formed. This view also agrees with the density estimates obtained from the inversions, which agree with the values obtained at the core of coronal rain clumps from simulations. From the distributions of density and temperature obtained from the spectral inversions, it seems that there is a noticeable change in the temperature and density of the plasma between the formation height of the $h_2$ and $k_2$ features, with the $h_2$ appearing to form in a slightly colder and denser region.

 The formation properties of the different spectral features of the  \ion{Mg}{II} h \& k lines have also allowed for a study of the velocity in the LoS direction. Several relationships obtained from the study of \citet{2013ApJ...772...90L} based in the comparison of synthetic line profiles obtained from 3D radiation transfer calculations in a snapshot of a 3D simulation were used in this study. If these relationships between spectral features and vertical velocities still hold for the case of the Sun in the presence of coronal rain clumps, the results of this work are suggestive of an interesting scenario at work. The plasma formed in the outer part of the coronal rain clump, where the line cores of the \ion{Mg}{II} h \& k lines are formed, falls towards the chromosphere with a greater speed than that of the plasma constituting the denser part of the clumps, where the emission peaks of the lines form. This situation seems to take place in all the observations, suggesting that it is a common phenomenon in the dynamics of coronal rain. It might not be present in all stages during the fall of the material and it might only occur in the initial part of the descent along magnetic field lines. The existence, cause, and possible significance of this velocity gradient should be analysed in future studies that would include numerical simulations and additional observations of on-disc coronal rain.  Some numerical simulations \citep[e.g.][]{2022ApJ...926L..29A, david2020} show that the falling velocity of the outer plasma layers is smaller than that of the inner parts. The interpretation of this fact is that less dense material falls more slowly than denser one \citep[see][]{oliver2014}. The results presented here, however, show that the material at the outer part of coronal rain falls faster than that closer to its core, where we can safely assume that the highest density is found. This discrepancy between numerical simulations and the present results warrants further attention in subsequent studies.

\begin{acknowledgements}
This publication is part of the R+D+i project PID2020-112791GB-I00, financed by MCIN/AEI/10.13039/501100011033. MK acknowledges the support from the Vicepresidència i Conselleria d’Innovació, Recerca i Turisme del Govern de les Illes Balears and the Fons Social Europeu 2014-2020 de les Illes Balears. IRIS is a NASA small explorer mission developed and operated by LMSAL with mission operations executed at NASA Ames Research center and major contributions to downlink communications funded by ESA and the Norwegian Space Centre. This research was supported by the International Space Science Institute (ISSI) in Bern, through ISSI International Team project number 545 (‘Observe Local Think Global: What Solar Observations can Teach us about Multiphase Plasmas across Physical Scales’). The authors thank the anonymous referee for some comments and suggestions that helped improve the manuscript.  
\end{acknowledgements}
\bibliographystyle{aa}
\bibliography{citations}

\begin{thebibliography}{39}
\expandafter\ifx\csname natexlab\endcsname\relax\def\natexlab#1{#1}\fi

\bibitem[{{Antiochos} \& {Klimchuk}(1991)}]{1991ApJ...378..372A}
{Antiochos}, S.~K. \& {Klimchuk}, J.~A. 1991, \apj, 378, 372

\bibitem[{{Antolin} {et~al.}(2022){Antolin}, {Mart{\'\i}nez-Sykora}, \&
  {{\c{S}}ahin}}]{2022ApJ...926L..29A}
{Antolin}, P., {Mart{\'\i}nez-Sykora}, J., \& {{\c{S}}ahin}, S. 2022, \apjl,
  926, L29

\bibitem[{{Antolin} \& {Rouppe van der Voort}(2012)}]{2012ApJ...745..152A}
{Antolin}, P. \& {Rouppe van der Voort}, L. 2012, \apj, 745, 152

\bibitem[{{Antolin} {et~al.}(2010){Antolin}, {Shibata}, \&
  {Vissers}}]{2010ApJ...716..154A}
{Antolin}, P., {Shibata}, K., \& {Vissers}, G. 2010, \apj, 716, 154

\bibitem[{{Antolin} {et~al.}(2015){Antolin}, {Vissers}, {Pereira}, {Rouppe van
  der Voort}, \& {Scullion}}]{2015ApJ...806...81A}
{Antolin}, P., {Vissers}, G., {Pereira}, T.~M.~D., {Rouppe van der Voort}, L.,
  \& {Scullion}, E. 2015, \apj, 806, 81

\bibitem[{{Avrett}(1985)}]{1985cdm..proc...67A}
{Avrett}, E.~H. 1985, in Chromospheric Diagnostics and Modelling, ed. B.~W.
  {Lites}, 67--127

\bibitem[{{Boerner} {et~al.}(2012){Boerner}, {Edwards}, {Lemen}, {Rausch},
  {Schrijver}, {Shine}, {Shing}, {Stern}, {Tarbell}, {Title}, {Wolfson},
  {Soufli}, {Spiller}, {Gullikson}, {McKenzie}, {Windt}, {Golub}, {Podgorski},
  {Testa}, \& {Weber}}]{2012SoPh..275...41B}
{Boerner}, P., {Edwards}, C., {Lemen}, J., {et~al.} 2012, \solphys, 275, 41

\bibitem[{{{\c{S}}ahin} {et~al.}(2023){{\c{S}}ahin}, {Antolin}, {Froment}, \&
  {Schad}}]{2023ApJ...950..171S}
{{\c{S}}ahin}, S., {Antolin}, P., {Froment}, C., \& {Schad}, T.~A. 2023, \apj,
  950, 171

\bibitem[{{da Silva Santos} {et~al.}(2018){da Silva Santos}, {de la Cruz
  Rodr{\'\i}guez}, \& {Leenaarts}}]{2018A&A...620A.124D}
{da Silva Santos}, J.~M., {de la Cruz Rodr{\'\i}guez}, J., \& {Leenaarts}, J.
  2018, \aap, 620, A124

\bibitem[{{de la Cruz Rodr{\'\i}guez} {et~al.}(2016){de la Cruz
  Rodr{\'\i}guez}, {Leenaarts}, \& {Asensio Ramos}}]{2016ApJ...830L..30D}
{de la Cruz Rodr{\'\i}guez}, J., {Leenaarts}, J., \& {Asensio Ramos}, A. 2016,
  \apjl, 830, L30

\bibitem[{{de la Cruz Rodr{\'\i}guez} {et~al.}(2019){de la Cruz
  Rodr{\'\i}guez}, {Leenaarts}, {Danilovic}, \&
  {Uitenbroek}}]{2019A&A...623A..74D}
{de la Cruz Rodr{\'\i}guez}, J., {Leenaarts}, J., {Danilovic}, S., \&
  {Uitenbroek}, H. 2019, \aap, 623, A74

\bibitem[{{de la Cruz Rodr{\'\i}guez} \&
  {Piskunov}(2013)}]{2013ApJ...764...33D}
{de la Cruz Rodr{\'\i}guez}, J. \& {Piskunov}, N. 2013, \apj, 764, 33

\bibitem[{{De Pontieu} {et~al.}(2014){De Pontieu}, {Title}, {Lemen}, {Kushner},
  {Akin}, {Allard}, {Berger}, {Boerner}, {Cheung}, {Chou}, {Drake}, {Duncan},
  {Freeland}, {Heyman}, {Hoffman}, {Hurlburt}, {Lindgren}, {Mathur}, {Rehse},
  {Sabolish}, {Seguin}, {Schrijver}, {Tarbell}, {W{\"u}lser}, {Wolfson},
  {Yanari}, {Mudge}, {Nguyen-Phuc}, {Timmons}, {van Bezooijen}, {Weingrod},
  {Brookner}, {Butcher}, {Dougherty}, {Eder}, {Knagenhjelm}, {Larsen},
  {Mansir}, {Phan}, {Boyle}, {Cheimets}, {DeLuca}, {Golub}, {Gates}, {Hertz},
  {McKillop}, {Park}, {Perry}, {Podgorski}, {Reeves}, {Saar}, {Testa}, {Tian},
  {Weber}, {Dunn}, {Eccles}, {Jaeggli}, {Kankelborg}, {Mashburn}, {Pust},
  {Springer}, {Carvalho}, {Kleint}, {Marmie}, {Mazmanian}, {Pereira}, {Sawyer},
  {Strong}, {Worden}, {Carlsson}, {Hansteen}, {Leenaarts}, {Wiesmann},
  {Aloise}, {Chu}, {Bush}, {Scherrer}, {Brekke}, {Martinez-Sykora}, {Lites},
  {McIntosh}, {Uitenbroek}, {Okamoto}, {Gummin}, {Auker}, {Jerram}, {Pool}, \&
  {Waltham}}]{2014SoPh..289.2733D}
{De Pontieu}, B., {Title}, A.~M., {Lemen}, J.~R., {et~al.} 2014, \solphys, 289,
  2733

\bibitem[{{Fontenla} {et~al.}(1993){Fontenla}, {Avrett}, \&
  {Loeser}}]{1993ApJ...406..319F}
{Fontenla}, J.~M., {Avrett}, E.~H., \& {Loeser}, R. 1993, \apj, 406, 319

\bibitem[{{Freeland} \& {Handy}(1998)}]{1998SoPh..182..497F}
{Freeland}, S.~L. \& {Handy}, B.~N. 1998, \solphys, 182, 497

\bibitem[{{Froment} {et~al.}(2020){Froment}, {Antolin}, {Henriques},
  {Kohutova}, \& {Rouppe van der Voort}}]{2020A&A...633A..11F}
{Froment}, C., {Antolin}, P., {Henriques}, V.~M.~J., {Kohutova}, P., \& {Rouppe
  van der Voort}, L.~H.~M. 2020, \aap, 633, A11

\bibitem[{{Gudiksen} {et~al.}(2011){Gudiksen}, {Carlsson}, {Hansteen}, {Hayek},
  {Leenaarts}, \& {Mart{\'\i}nez-Sykora}}]{2011A&A...531A.154G}
{Gudiksen}, B.~V., {Carlsson}, M., {Hansteen}, V.~H., {et~al.} 2011, \aap, 531,
  A154

\bibitem[{{Kriginsky} {et~al.}(2021){Kriginsky}, {Oliver}, {Antolin},
  {Kuridze}, \& {Freij}}]{2021A&A...650A..71K}
{Kriginsky}, M., {Oliver}, R., {Antolin}, P., {Kuridze}, D., \& {Freij}, N.
  2021, \aap, 650, A71

\bibitem[{{Kriginsky} {et~al.}(2023){Kriginsky}, {Oliver}, \&
  {Kuridze}}]{2023A&A...672A..89K}
{Kriginsky}, M., {Oliver}, R., \& {Kuridze}, D. 2023, \aap, 672, A89

\bibitem[{{Leenaarts} \& {Carlsson}(2009)}]{2009ASPC..415...87L}
{Leenaarts}, J. \& {Carlsson}, M. 2009, in Astronomical Society of the Pacific
  Conference Series, Vol. 415, The Second Hinode Science Meeting: Beyond
  Discovery-Toward Understanding, ed. B.~{Lites}, M.~{Cheung}, T.~{Magara},
  J.~{Mariska}, \& K.~{Reeves}, 87

\bibitem[{{Leenaarts} {et~al.}(2007){Leenaarts}, {Carlsson}, {Hansteen}, \&
  {Rutten}}]{2007A&A...473..625L}
{Leenaarts}, J., {Carlsson}, M., {Hansteen}, V., \& {Rutten}, R.~J. 2007, \aap,
  473, 625

\bibitem[{{Leenaarts} {et~al.}(2012){Leenaarts}, {Pereira}, \&
  {Uitenbroek}}]{2012A&A...543A.109L}
{Leenaarts}, J., {Pereira}, T., \& {Uitenbroek}, H. 2012, \aap, 543, A109

\bibitem[{{Leenaarts} {et~al.}(2013{\natexlab{a}}){Leenaarts}, {Pereira},
  {Carlsson}, {Uitenbroek}, \& {De Pontieu}}]{2013ApJ...772...89L}
{Leenaarts}, J., {Pereira}, T.~M.~D., {Carlsson}, M., {Uitenbroek}, H., \& {De
  Pontieu}, B. 2013{\natexlab{a}}, \apj, 772, 89

\bibitem[{{Leenaarts} {et~al.}(2013{\natexlab{b}}){Leenaarts}, {Pereira},
  {Carlsson}, {Uitenbroek}, \& {De Pontieu}}]{2013ApJ...772...90L}
{Leenaarts}, J., {Pereira}, T.~M.~D., {Carlsson}, M., {Uitenbroek}, H., \& {De
  Pontieu}, B. 2013{\natexlab{b}}, \apj, 772, 90

\bibitem[{{Lemen} {et~al.}(2012){Lemen}, {Title}, {Akin}, {Boerner}, {Chou},
  {Drake}, {Duncan}, {Edwards}, {Friedlaender}, {Heyman}, {Hurlburt}, {Katz},
  {Kushner}, {Levay}, {Lindgren}, {Mathur}, {McFeaters}, {Mitchell}, {Rehse},
  {Schrijver}, {Springer}, {Stern}, {Tarbell}, {Wuelser}, {Wolfson}, {Yanari},
  {Bookbinder}, {Cheimets}, {Caldwell}, {Deluca}, {Gates}, {Golub}, {Park},
  {Podgorski}, {Bush}, {Scherrer}, {Gummin}, {Smith}, {Auker}, {Jerram},
  {Pool}, {Soufli}, {Windt}, {Beardsley}, {Clapp}, {Lang}, \&
  {Waltham}}]{2012SoPh..275...17L}
{Lemen}, J.~R., {Title}, A.~M., {Akin}, D.~J., {et~al.} 2012, \solphys, 275, 17

\bibitem[{{Mart{\'\i}nez-G{\'o}mez} {et~al.}(2020){Mart{\'\i}nez-G{\'o}mez},
  {Oliver}, {Khomenko}, \& {Collados}}]{david2020}
{Mart{\'\i}nez-G{\'o}mez}, D., {Oliver}, R., {Khomenko}, E., \& {Collados}, M.
  2020, \aap, 634, A36

\bibitem[{{Miki{\'c}} {et~al.}(2013){Miki{\'c}}, {Lionello}, {Mok}, {Linker},
  \& {Winebarger}}]{2013ApJ...773...94M}
{Miki{\'c}}, Z., {Lionello}, R., {Mok}, Y., {Linker}, J.~A., \& {Winebarger},
  A.~R. 2013, \apj, 773, 94

\bibitem[{{M{\"u}ller} {et~al.}(2003){M{\"u}ller}, {Hansteen}, \&
  {Peter}}]{2003A&A...411..605M}
{M{\"u}ller}, D.~A.~N., {Hansteen}, V.~H., \& {Peter}, H. 2003, \aap, 411, 605

\bibitem[{{M{\"u}ller} {et~al.}(2004){M{\"u}ller}, {Peter}, \&
  {Hansteen}}]{2004A&A...424..289M}
{M{\"u}ller}, D.~A.~N., {Peter}, H., \& {Hansteen}, V.~H. 2004, \aap, 424, 289

\bibitem[{{Oliver} {et~al.}(2014){Oliver}, {Soler}, {Terradas}, {Zaqarashvili},
  \& {Khodachenko}}]{oliver2014}
{Oliver}, R., {Soler}, R., {Terradas}, J., {Zaqarashvili}, T.~V., \&
  {Khodachenko}, M.~L. 2014, \apj, 784, 21

\bibitem[{{Pereira} {et~al.}(2015){Pereira}, {Carlsson}, {De Pontieu}, \&
  {Hansteen}}]{2015ApJ...806...14P}
{Pereira}, T. M.~D., {Carlsson}, M., {De Pontieu}, B., \& {Hansteen}, V. 2015,
  \apj, 806, 14

\bibitem[{{Pesnell} {et~al.}(2012){Pesnell}, {Thompson}, \&
  {Chamberlin}}]{2012SoPh..275....3P}
{Pesnell}, W.~D., {Thompson}, B.~J., \& {Chamberlin}, P.~C. 2012, \solphys,
  275, 3

\bibitem[{{Piskunov} \& {Valenti}(2017)}]{2017A&A...597A..16P}
{Piskunov}, N. \& {Valenti}, J.~A. 2017, \aap, 597, A16

\bibitem[{{Scharmer} {et~al.}(2003){Scharmer}, {Bjelksjo}, {Korhonen},
  {Lindberg}, \& {Petterson}}]{sst}
{Scharmer}, G.~B., {Bjelksjo}, K., {Korhonen}, T.~K., {Lindberg}, B., \&
  {Petterson}, B. 2003, 4853

\bibitem[{Scharmer {et~al.}(2008)Scharmer, Narayan, Hillberg, de~la
  Cruz~Rodriguez, Löfdahl, Kiselman, Sütterlin, van Noort, \&
  Lagg}]{Scharmer_2008}
Scharmer, G.~B., Narayan, G., Hillberg, T., {et~al.} 2008, The Astrophysical
  Journal, 689, L69

\bibitem[{{Tsuneta} {et~al.}(2008){Tsuneta}, {Ichimoto}, {Katsukawa}, {Nagata},
  {Otsubo}, {Shimizu}, {Suematsu}, {Nakagiri}, {Noguchi}, {Tarbell}, {Title},
  {Shine}, {Rosenberg}, {Hoffmann}, {Jurcevich}, {Kushner}, {Levay}, {Lites},
  {Elmore}, {Matsushita}, {Kawaguchi}, {Saito}, {Mikami}, {Hill}, \&
  {Owens}}]{2008SoPh..249..167T}
{Tsuneta}, S., {Ichimoto}, K., {Katsukawa}, Y., {et~al.} 2008, \solphys, 249,
  167

\bibitem[{{Uitenbroek}(2001)}]{2001ApJ...557..389U}
{Uitenbroek}, H. 2001, \apj, 557, 389

\bibitem[{{Vissers} {et~al.}(2019){Vissers}, {de la Cruz Rodr{\'\i}guez},
  {Libbrecht}, {Rouppe van der Voort}, {Scharmer}, \&
  {Carlsson}}]{2019A&A...627A.101V}
{Vissers}, G.~J.~M., {de la Cruz Rodr{\'\i}guez}, J., {Libbrecht}, T., {et~al.}
  2019, \aap, 627, A101

\bibitem[{{W{\"u}lser} {et~al.}(2018){W{\"u}lser}, {Jaeggli}, {De Pontieu},
  {Tarbell}, {Boerner}, {Freeland}, {Liu}, {Timmons}, {Brannon}, {Kankelborg},
  {Madsen}, {McKillop}, {Prchlik}, {Saar}, {Schanche}, {Testa}, {Bryans}, \&
  {Wiesmann}}]{2018SoPh..293..149W}
{W{\"u}lser}, J.~P., {Jaeggli}, S., {De Pontieu}, B., {et~al.} 2018, \solphys,
  293, 149

\end{thebibliography}
\end{document}